\documentclass[journal=nalefd,manuscript=article, layout=onecolumn]{achemso}
\setkeys{acs}{articletitle = true}

\usepackage[version=3]{mhchem} 
\usepackage[utf8]{inputenc} 

\usepackage{xcolor}
\newcommand{\comment}[1]{ {\bfseries \color{blue} XXX #1 XXX}} 

\usepackage{ulem}

\usepackage{lineno}

\definecolor{rv}{rgb}{1, 0, 0}

\author{Guillaume Sarfati}
\affiliation[]{\small Sorbonne Universit\'e, CNRS, Institut de Biologie Paris-Seine (IBPS), Laboratoire Jean Perrin (LJP), F-75005, Paris}

\author{Ananyo Maitra}
\affiliation[]{\small Laboratoire de Physique Th\'eorique et Mod\'elisation, CNRS UMR 8089,
	CY Cergy Paris Universit\'e, F-95302 Cergy-Pontoise Cedex, France}

\author{Raphael Voituriez}
\affiliation[]{\small Sorbonne Universit\'e, CNRS, Institut de Biologie Paris-Seine (IBPS), Laboratoire Jean Perrin (LJP), F-75005, Paris}
\alsoaffiliation{Sorbonne Universit\'e, CNRS, Laboratoire de Physique Théorique de la Matière Condensée (LPTMC), F-75005 Paris, France}

\author{Jean-Christophe Galas}
\affiliation[]{\small Sorbonne Universit\'e, CNRS, Institut de Biologie Paris-Seine (IBPS), Laboratoire Jean Perrin (LJP), F-75005, Paris}
\email{jean-christophe.galas@sorbonne-universite.fr}

\author{Andr\'e Estevez-Torres}
\affiliation[]{\small Sorbonne Universit\'e, CNRS, Institut de Biologie Paris-Seine (IBPS), Laboratoire Jean Perrin (LJP), F-75005, Paris}
\email{andre.estevez-torres@sorbonne-universite.fr}

\title[]{
Crosslinking and depletion determine spatial instabilities in cytoskeletal active matter
}%
\date{today}

\usepackage{amsmath}
\usepackage{amssymb}
\usepackage{lipsum}
\usepackage{bm}
\usepackage{xcolor}

\usepackage{graphicx}
\usepackage[T1]{fontenc}
\usepackage{subcaption}
\usepackage{hyperref}

\def\be{\begin{equation}}
	\def\ee{\end{equation}}
\def\bea{\begin{eqnarray}}
	\def\eea{\end{eqnarray}}

\def\besub{\begin{subequations}}
	\def\eesub{\end{subequations}}

\def\bwd{\begin{widetext}}
	\def\ewd{\end{widetext}}

\definecolor{ao(english)}{rgb}{0.0, 0.5, 0.0}
\definecolor{armygreen}{rgb}{0.29, 0.33, 0.13}
\definecolor{auburn}{rgb}{0.43, 0.21, 0.1}
\definecolor{brightmaroon}{rgb}{0.76, 0.13, 0.28}
\definecolor{cadmiumred}{rgb}{0.89, 0.0, 0.13}
\definecolor{carnelian}{rgb}{0.7, 0.11, 0.11}
\definecolor{cornellred}{rgb}{0.7, 0.11, 0.11}
\definecolor{crimsonglory}{rgb}{0.75, 0.0, 0.2}
\definecolor{orangeyellow}{rgb}{0.3, 0.2, 0.2}
\definecolor{fluorescentorange}{rgb}{1.0, 0.75, 0.0}
\definecolor{gamboge}{rgb}{0.89, 0.61, 0.06}
\newcommand{\bsf}[1]{\textsf{\textbf{#1}}}

\begin{document}

\maketitle

\begin{abstract}
	Active gels made of cytoskeletal proteins are valuable materials with attractive non-equilibrium properties such as spatial self-organization and self-propulsion. At least four typical routes to spatial patterning have been reported to date in different types of cytoskeletal active gels: bending and buckling instabilities in extensile systems, and global and local contraction instabilities in contractile gels. Here we report the observation of these four instabilities in a single type of active gel and we show that they are controlled by two parameters: the concentrations of ATP and depletion agent. We demonstrate that as the ATP concentration decreases, the concentration of passive motors increases until the gel undergoes a gelation transition. At this point, buckling is selected against bending, while global contraction is favored over local ones. Our observations are coherent with a hydrodynamic model of a viscoelastic active gel where the filaments are crosslinked with a characteristic time that diverges as the ATP concentration decreases. Our work thus provides a unified view of spatial instabilities in cytoskeletal active matter.
	\end{abstract}


\section{Introduction}

Active matter is composed of agents that transform free energy into mechanical work, thus making non-equilibrium materials with attractive properties such as spatial self-organization and self-propulsion\cite{marchetti2013hydrodynamics, needleman_active_2017, bricard2013}. In this context, cytoskeletal active matter, constituted of protein filaments set in motion by molecular motors, is particularly interesting because the collective interaction of nanometric subunits creates order at the macroscopic scale. However, the determinants of spatial self-organization in these active systems are not yet well understood. 3-dimensional cytoskeletal active matter is prone to at least four different spatial instabilities: i) bending\cite{chandrakar_confinement_2020}, ii) buckling\cite{senoussi_tunable_2019, strubing_wrinkling_2020, ideses2018}, iii) global\cite{bendix_2008, torisawa_spontaneous_2016, alvarado_molecular_2013, foster_active_2015} and iv) local contractions\cite{nedelec_self-organization_1997, alvarado_molecular_2013, torisawa_spontaneous_2016}. 
These instabilities have been observed experimentally but so far in different systems, thus precluding a unified analysis of potential transitions between all possible phases. Further, there is currently no clear theoretical understanding of how a gel transitions from one instability mode to another.

Here we consider a mixture composed of microtubules (MTs), kinesin motor clusters that pull on them and a depletant that generates microtubule bundles\cite{sanchez_spontaneous_2012}. In the presence of ATP, this mixture generates an active solution in 3-dimensions that we will call `Dogic system' in the following. When the MTs are stabilized with GMPCPP they are shorter and the Dogic system generates continuous chaotic flows\cite{sanchez_spontaneous_2012} through an in-plane bending instability both in 2D\cite{martinez2019} and   in 3D\cite{chandrakar_confinement_2020}. When the MTs are stabilized with taxol they are longer and chaotic flows are still observed but they emerge only after an out-of-plane buckling instability that creates a corrugated sheet\cite{senoussi_tunable_2019, strubing_wrinkling_2020}. Bending\cite{kumar2018} and buckling\cite{ideses2018} have also been reported in actin-myosin gels.

Besides bending and buckling instabilities, global and local contractions have been reported both in actin-myosin gels\cite{bendix_2008, alvarado_molecular_2013} and in MT-motors mixtures\cite{foster_active_2015, torisawa_spontaneous_2016, nedelec_self-organization_1997, surrey_physical_2001}. 
In the Dogic system, global contractions appear with taxol MTs \cite{strubing_wrinkling_2020, nasirimarekani_tuning_2021, senoussi_biorxiv_2021, senoussi_sciadv_2021}, while local contractions arise with GMPCPP MTs \cite{senoussi_biorxiv_2021, senoussi_sciadv_2021}. Very recent experiments, however, have observed both types of contractions with GMPCPP MTs\cite{lemma_active_2021}. Finally, dynamic microtubules in the presence of multiheaded-kinesins yield chaotic flows or local contractions in the absence of depletant\cite{roostalu_determinants_2018} depending on the ratio between MT growth and motor speed, and the ratio between the motor and MT concentration. In summary, the disparity of conditions in which spatial instabilities have been observed in cytoskeletal active matter (Tab.~S\ref{Tab_SI_biblio}) hinders our understanding of these systems. Here, we demonstrate that the four instabilities can be observed in the Dogic system by tuning only two parameters: ATP and depletant concentrations. The transition from bending to buckling and from local to global contractions happens at a critical ATP concentration that is associated to a critical concentration of passive motors that induces a gelation transition between a liquid-like and a solid-like state\cite{gagnon_shear-induced_2020}. A similar transition is observed at constant ATP concentration when the concentration of a passive linker is increased, which supports its generic physical origin, independently of the specific nature of the crosslinking agents.

\section{Results and discussion}
\begin{figure}
	\includegraphics[width=\linewidth]{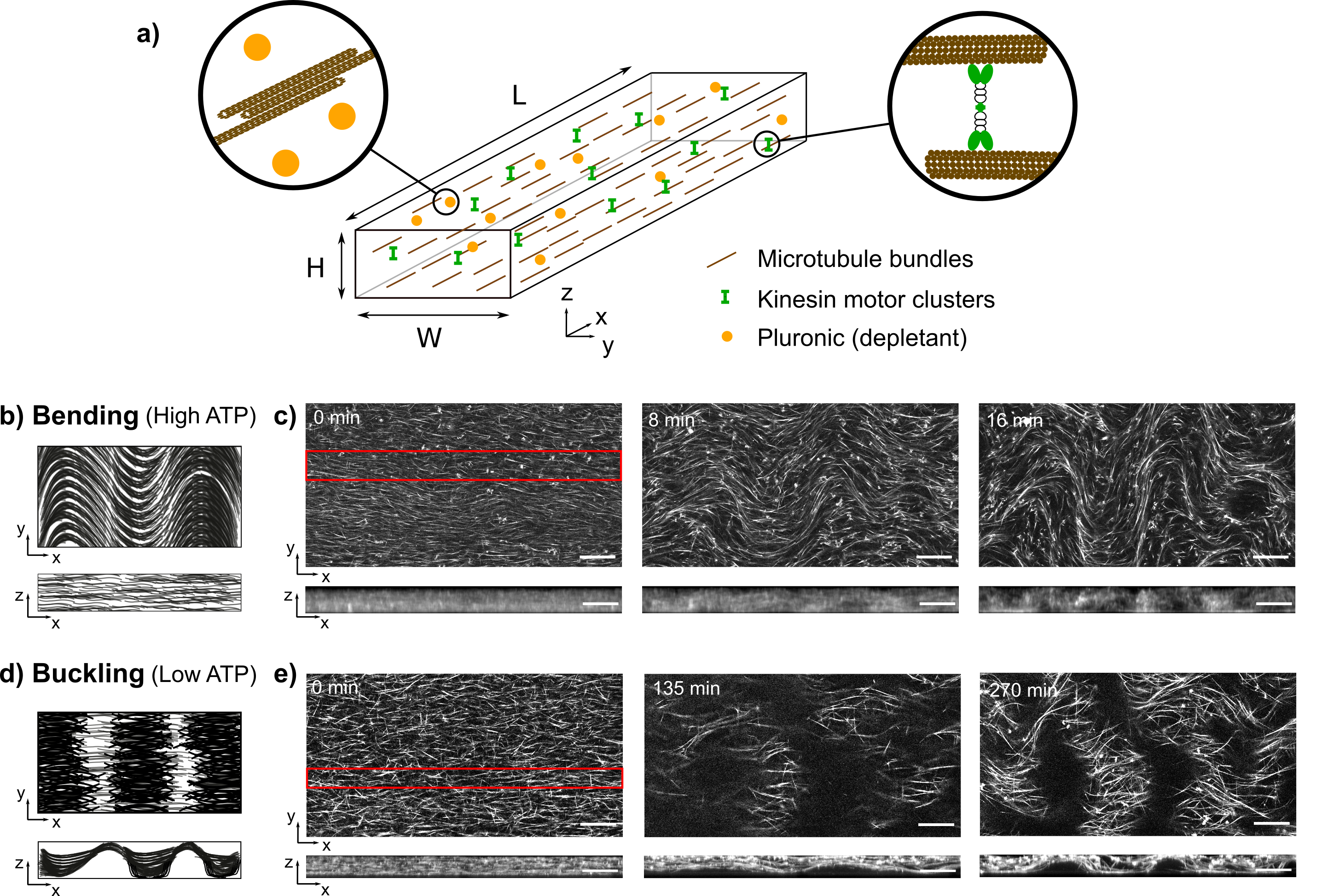}
	\caption{Bending and buckling instabilities are controlled by ATP concentration at high depletant concentration.  a. Scheme of the experimental system with the microtubules in brown, the motor clusters in green and the pluronic depletant in orange, inside a rectangular channel. Cartoons of the microtubule morphology during the bending (b) and buckling (d) instabilities (the MTs appear in black). Time-lapse confocal images of fluorescent microtubules showing the bending (c) and buckling (e) from the bottom ($xy$-plane) and from the side ($xz$-plane) of the channel (the MTs appear in white). 5.5~\% pluronic, [ATP] = 50~$\mu$M for bending and  5~$\mu$M for buckling. Scale bars are 200~$\mu$m.}
	\label{fig:two_instabilities}
\end{figure}

Our experiments take place in a long and thin channel of rectangular cross-section with typical dimensions $L\times W \times H = 10\times1\times0.1$ mm$^3$. The active solution is composed of short GMPCPP-stabilized microtubules, clusters of biotinylated kinesin-1 in the presence of streptavidin, the depletant agent pluronic, that forms MTs bundles, and ATP coupled to an ATP-regeneration system (see Fig.~\ref{fig:two_instabilities}a and SI Methods). MTs are fluorescent and they are imaged using either confocal or epifluorescence microscopy. When the solution is pipetted into the channel, the shear flow generates an initial nematic order along the $x$-axis. 

At high pluronic concentration (5.5~\% w/v), the nematic state is deformed in two different ways depending on the ATP concentration, before chaotic flow occurs. At 50~$\mu$M ATP, the MT bundles deform in the $xy$-plane forming waves with a typical wavelength of 250~$\mu$m along the $x$-axis that arises within 5 minutes  (Fig \ref{fig:two_instabilities}b,c and Movie~S\ref{Mov_Fig1_Bending}). Confocal images show a minimal deformation of the bundles in the $xz$-plane. This in-plane bending instability has recently been observed in a similar system both in 3D\cite{chandrakar_confinement_2020} and in 2D\cite{martinez2019}. Within $10-20$~min, a superposition of bending instabilities results in a chaotic flow of bundles (Movie~S\ref{Mov_Bending_TempsLong}). At  5~$\mu$M ATP, in contrast, the MTs deform in the $xz$-plane, creating an undulating sheet with a typical wavelength of 200~$\mu$m within 150~min (Fig.~\ref{fig:two_instabilities}d,e and Movie~S\ref{Mov_Fig1_Buckling}). In the $xy$-plane, this sheet appears as a periodic structure of black and bright zones, with preserved nematic order of MT bundles along the  $x$-axis. Within 1500~min this instability conduces to an onset of chaotic flows destroying nematic order, although the slow dynamics preclude the observation of a fully developed chaotic state (Movie~S\ref{Mov_Buckling_TempsLong}). This out-of-plane buckling instability has recently been reported in a system that is similar but displays two important differences: longer, taxol-stabilized MTs, and much higher [ATP]  (2 mM)\cite{senoussi_tunable_2019, strubing_wrinkling_2020}.

\begin{figure}
	\includegraphics[width=\linewidth]{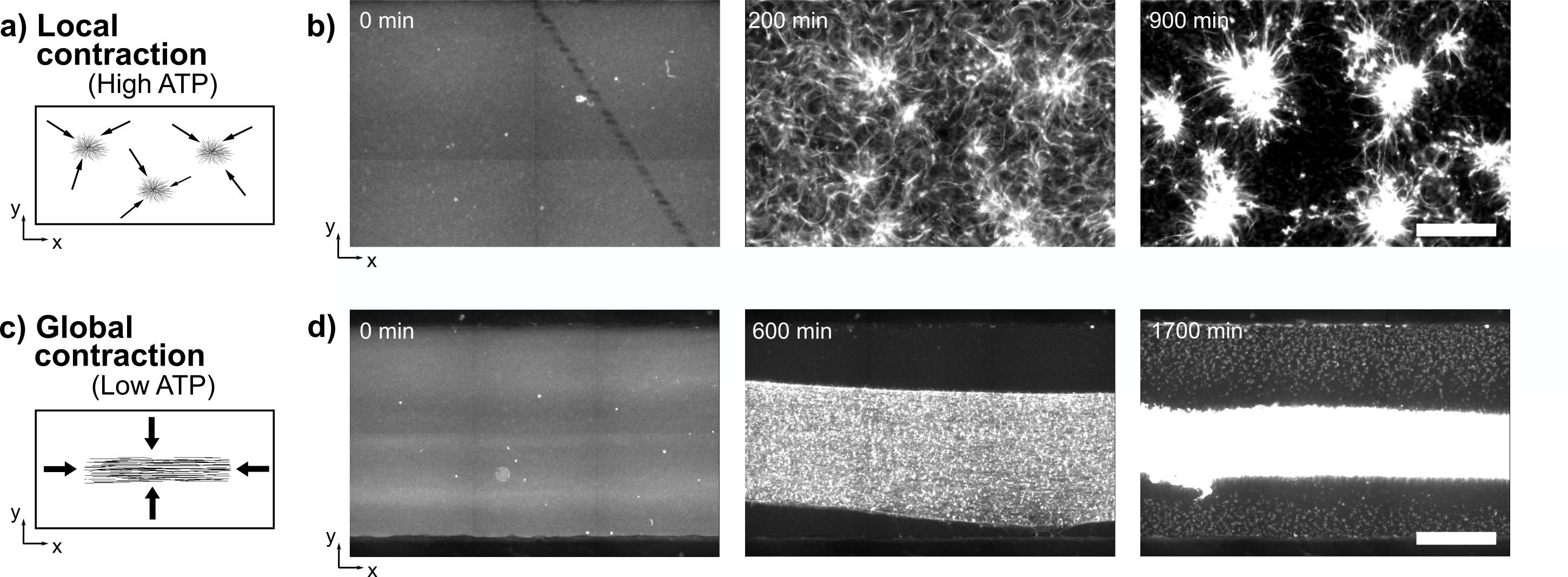}
	\caption{Global and local contractions are controlled by ATP concentration at low depletant concentration. 
	Cartoons of the microtubule morphology during local (a) and global (c) contractions (the MTs appear in black). Time-lapse epifluorescence images for local (b) and global contractions (d) in the $yx$-plane (the MTs appear in white). 1.5~\% pluronic, [ATP] = 5~$\mu$M for global and  100~$\mu$M for local contractions. Scale bars are 500~$\mu$m.}
	\label{fig:contractions}
\end{figure}

At a lower pluronic concentration (1.5~\%) the behavior of the system changes dramatically (Fig.~\ref{fig:contractions}).  In these conditions and high [ATP] (100~$\mu$M), a chaotic flow is observed during the first 100 min but afterwards microtubules aggregate, their area growing with an exponential rate of $0.028~$min$^{-1}$, first to make 20~$\mu$m diameter clumps that coalesce after 1500 min into $\sim250~\mu$m diameter aggregates separated by a typical distance of  $\sim500~\mu$m (Movie~S\ref{Mov_Fig1_LocalContraction}). We call this late instability local contractions. At the time resolution of the experiments, the type of instability that yields the initial chaotic flow could not be resolved. In contrast, at low [ATP]  (5~$\mu$M), a global contraction in the absence of chaotic flow is observed (Movie~S\ref{Mov_Fig1_GlobalContraction}). The MT network first contracts along $y$ with an onset rate of $0.015~$min$^{-1}$ until it reaches a steady-state width of $0.3W$ at $t=1000$~min and later contracts along $x$ 
until contraction stops at 1700~min. This transition from local to global contraction has recently been found to be strongly dependent on MT concentration, slightly dependent on depletant concentration and independent of motor concentration in the Dogic system\cite{lemma_active_2021}. In addition, it reminds the one reported for actomyosin gels, where a critical concentration of passive linkers induced the percolation of the gel yielding the global contraction state\cite{alvarado_molecular_2013}. 

\begin{figure}
	\includegraphics[width=0.7\linewidth]{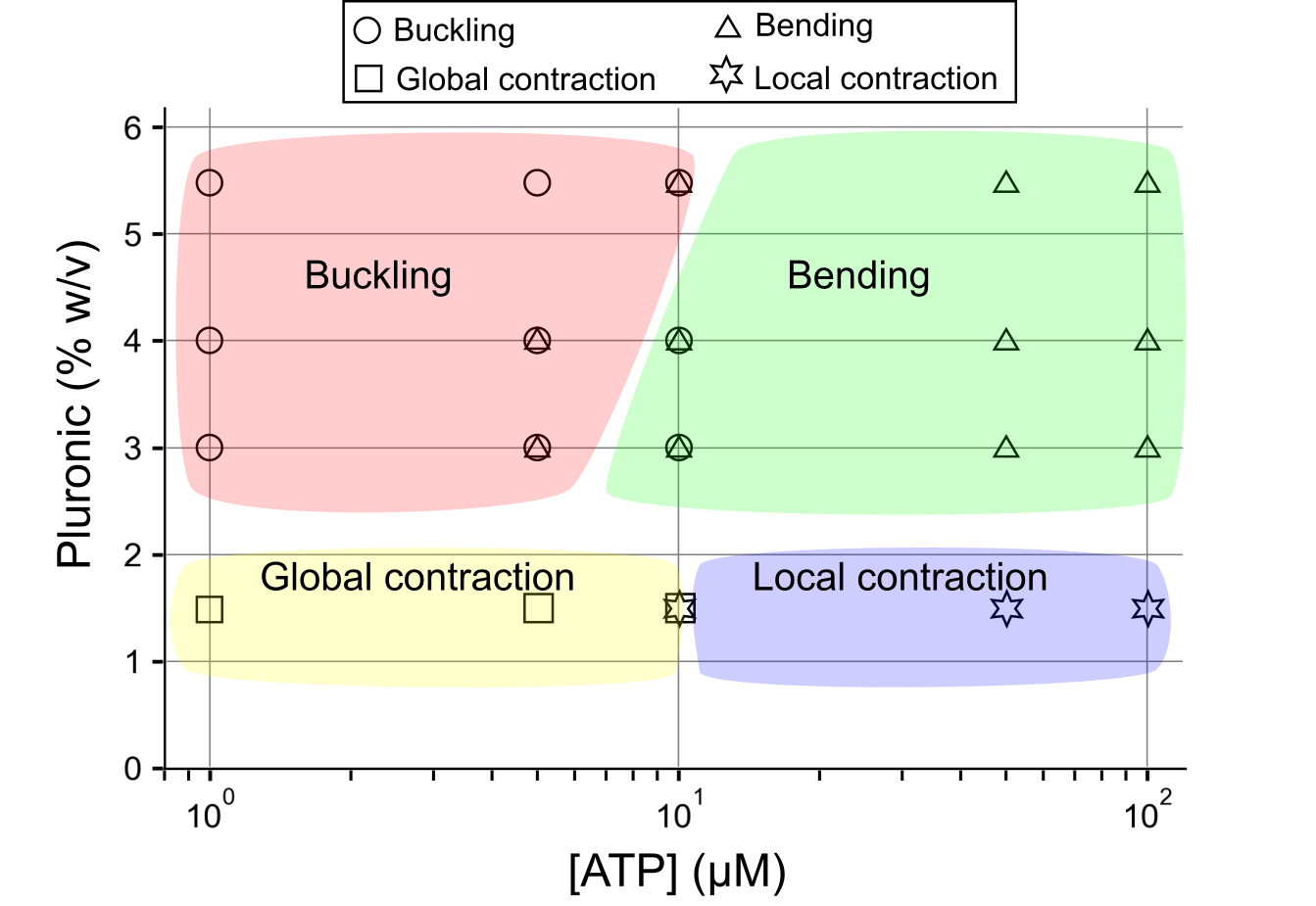}
	\caption{Phase space of the four spatial instabilities as a function of pluronic and ATP concentrations. The symbols indicate the observed instabilities in a triplicate experiment and the colors are a guide to the eye.}
	\label{PhaseSpace}
\end{figure}

Fig. \ref{PhaseSpace} recapitulates the phase diagram of spatial instabilities as a function of depletant and ATP concentrations. The transition between bending and buckling happens at a critical concentration $[\textrm{ATP}]_c$ in the range $5-10~\mu$M. When the pluronic concentration increases, the $[\textrm{ATP}]_c$ marginally increases. 
Note that the characteristic time of the instability strongly decreases with increasing [ATP] (Fig.~S\ref{FigSI_BendBuck_overTime} and Table~S\ref{si_tab_char_time}), with buckling being significantly slower than bending, except at $[\textrm{ATP}]_c$ where replicate experiments yield either one or the other instability or a mixture of the two (Fig.~S\ref{FigSI_transition_bending_buckling}), with identical time-scales. 

The transition between local and global contractions also happens at $[\textrm{ATP}]_c =10~\mu$M, suggesting that ATP plays a similar role in both processes. Interestingly, for long, taxol-stabilized MTs a similar phase diagram was observed, although with some differences (Fig.~S\ref{FigSI_PhaseSpaceTaxol}). First, the bending/buckling and local/global transition happened at a slightly higher ATP concentration. Second, the limit between contracting and extensile instabilities occurred at a lower depletant concentration. Finally, in some occasions, buckling and bending ended up producing, respectively, global and local contractions instead of chaotic flows.

\subsection{The bending/buckling transition is not controlled by passive contraction}

\begin{figure}
	\includegraphics[width=\linewidth]{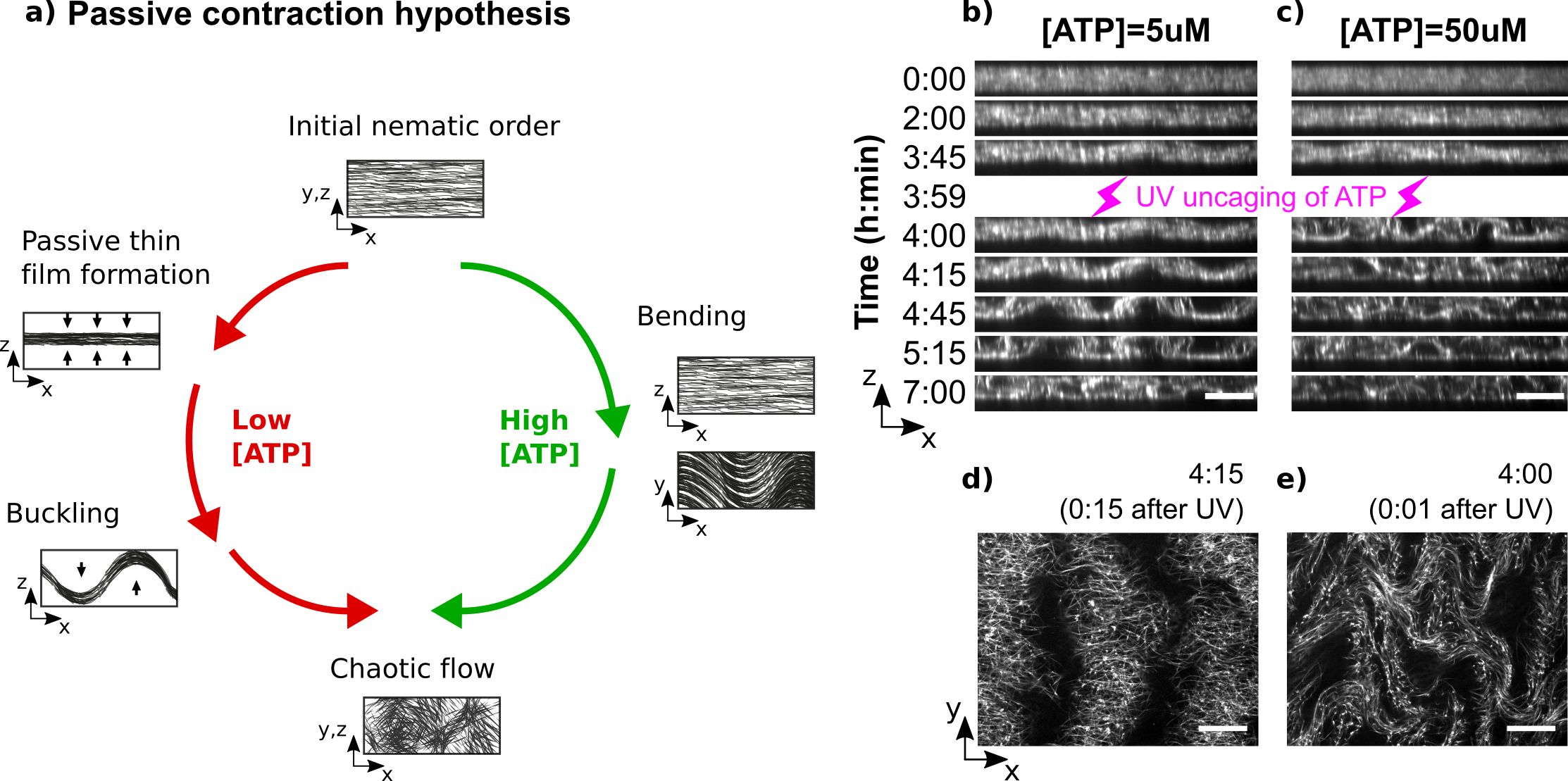}
	\caption{The bending/buckling transition is not controlled by passive contraction. a) Scheme of the kinetic hypothesis, where
	two instabilities are observed depending on ATP concentration.
	At low [ATP], activity is reduced, which lets the time for the passive contraction of microtubules to form a film that buckles.
	At high [ATP], the higher activity induces a bending instability.
 Confocal images of solutions with respectively 5~$\mu$M (b, d) and 50~$\mu$M  (c, e) caged-ATP.  Panels b and c provide time-lapses in the $xz$-plane with the uncaging event indicated at 3~h 59~min. Panels d and e display images in the $xy$-plane just after uncaging. Scale bars are 100~$\mu$m in $xy$ and 188~$\mu$m in $z$. Pluronic 5.5~\% w/v.}
	\label{fig:cagedATP}
\end{figure}

Senoussi et al. and Str\"ubing et al. observed a buckling instability at high [ATP] in the presence of long MTs.\cite{senoussi_tunable_2019, strubing_wrinkling_2020}. In ref.~\citenum{senoussi_tunable_2019}, the MTs contracted along $z$ due to passive depletion forces to form a film that buckled because of the negative surface tension induced by the motors. Our first hypothesis is thus that the reduction in [ATP] reduces the activity and lets the time for the passive contraction to form a film that  buckles. In this kinetic hypothesis two processes are in competition: the passive contraction of MTs into a film, whose time-scale would be independent of [ATP], and the activity due to the motors, whose dynamics may depend on [ATP] (Fig.~\ref{fig:cagedATP}a). To test it we first measured the passive contraction dynamics along $z$ in the absence of ATP (Figure~S\ref{FigSI_PassiveContraction}). In the absence of pluronic but in the presence of motors no contraction was observed. The strongest contraction, 50 \% of the thickness of the gel, was observed in the presence of pluronic alone and it reached 30 \% when both kinesin and pluronic were present. The characteristic time of passive contraction in these two cases was  in the range $150-200$~min (Fig.~S\ref{FigSI_PassiveContraction}). 

We then repeated the bending/buckling experiment in Fig.~\ref{fig:two_instabilities} but using caged-ATP to let the passive contraction happen in the absence of activity. Initially, the two systems at 5 and $50~\mu$M caged-ATP evolved in the absence of free ATP and after 4~h a light pulse uncaged it (Fig.~\ref{fig:cagedATP}). In both cases, MTs  contracted slightly along $z$ during the first 4~h (Fig.~S\ref{FigSI_PassiveContraction}b). After uncaging, buckling with preserved nematic order along $x$ was observed at $5~\mu$M ATP within 15 min (Fig.~\ref{fig:cagedATP}b,d). At  $50~\mu$M ATP, uncaging resulted in a bending instability destroying nematic order along $x$ (Fig.~\ref{fig:cagedATP}c,e), as in happened in Fig.~\ref{fig:two_instabilities}c, although the bending was less clear. Indeed, this is expected because, as discussed in the following, a suspended film that bends in the $xy$-plane necessarily deforms in the $xz$-plane. Importantly, however, pure buckling was not observed at $50~\mu$M ATP even though the uncaging occurred after the MTs film was formed. This result thus invalidates the kinetic hypothesis.

\subsection{A passive crosslinker-based theory to explain the transition from one instability mode to another}

Gagnon et al. used micro-rheology experiments to show that reducing  [ATP] in the Dogic system results in a gelation transition from a fluid to a solid-like state\cite{gagnon_shear-induced_2020} because kinesin motors act as long-lived passive linkers in the absence of ATP. Our second hypothesis is thus that the motors, in the low ATP regime,  control the viscoelastic properties of the system, thereby changing  the type of instability  that is observed, from in-plane bending in the fluid phase at high [ATP] to out-of-plane buckling in the solid phase at low [ATP] (Fig.~\ref{fig:PRC1}a,b). In this subsection, we use a hydrodynamic active matter framework \cite{RMP, SR_rev, Prost2015, SalJul_RepPhys} to show that the observation of different instability modes is indeed consistent with a gelation transition controlled by crosslinking.

 We model the kinesin-microtubule gel using a consistent description of an orientable and crosslinked active material described in ref.~\citenum{Ano_poly} ; this general theory is reminded in the SI and summarized in the following. We assume that at low [ATP] most kinesin motors act as passive crosslinkers, while at high [ATP] most motors are activated and therefore do not act as passive crosslinkers. The concentrations of passive or active motors do not appear directly in our theory; instead we start from a classical  model of viscoelasticity \cite{Beris1994} where crosslinkers effectively control the relaxation rate $1/\tau_C$ of a conformation tensor ${\bsf C}$, which characterizes the polymeric conformation of the microtubule filaments.   $\tau_C\to 0$  corresponds to a fluid state (at high  [ATP]), while $\tau_C\to\infty$ corresponds to a solid state (at low [ATP]). In such a permanently crosslinked, solid regime, the deviation of the conformation tensor, $\delta {\bsf C}$, from its steady-state value is the classical strain tensor of the gel \cite{Ano_poly}. Next, the degree of alignment of the microtubules is described by the nematic  tensor ${\bsf Q}$, as is usual in the context of liquid crystals.  Last, the dynamics of the gel is characterized by the fluid flow ${\bf v}$. The general hydrodynamic equations governing the dynamics of the coupled variables $\bsf C$, $\bsf Q$, ${\bf v}$ are displayed in the SI (Eqs. \ref{eq_tensorQ}, \ref{eq_tensorC} and \ref{eqvel}).  In particular, they include a free energetic coupling between alignment $\bsf Q$ and conformation $\bsf C$ of the gel with a coefficient $\chi$ (SI Eq.~\ref{fenrg1}), as well as  couplings of both ${\bsf C}$ and ${\bsf Q}$ to the fluid flow ${\bf v}$. Activity is introduced in the model via an active stress $\propto\zeta{\bsf Q}$ that only depends on the orientation, with $\zeta>0$ describing an extensile system (SI Eq.~\ref{eqstrs}).
 
This model provides four major results. First, in the fluid limit ($\tau_C\to 0$, associated to high [ATP]),  and for high enough active drive $\zeta$, the classical  in-plane bending instability characteristic of extensile fluids \cite{RMP, Voit, Aditi1} is expected (SI Eq.~\ref{eq_fluid_bending_inst}). Second, in the solid limit ($\tau_C\to \infty$, associated to low [ATP]), the bending instability is suppressed when $\chi>0$ (SI Eq.~\ref{eq_solid_no_bending_inst}). Third, in the solid limit, if the gel disposes of some space to deform in the $z$ direction, it will buckle out of plane because the extensile active stress $\zeta{\bsf Q}$ leads to a {\it negative} and destabilising surface tension  along the uniaxial direction $x$, as previously shown\cite{senoussi_tunable_2019} (SI Eq.~\ref{eq_solid_buckling_inst}). Importantly, a buckling solid will not bend  (at least initially). In contrast, the fourth result shows that a fluid that may deform in the $z$ direction will show both bending and buckling instabilities (SI Eq.~\ref{eq_fluid_film_bending_and_buckling_inst}). These theoretical results are compatible with the previous experimental observations if we suppose, based on Gagnon et al\cite{gagnon_shear-induced_2020}, that [ATP] drives a gelation transition in the Dogic system. In particular, the third and fourth theoretical results are consistent with buckling being observed at low [ATP] and a mixture of buckling and bending at high [ATP] when the active system may deform in the $z$ axis, as shown in the experiments in Fig.~\ref{fig:cagedATP}.

Note that while we assumed that passive motors act as crosslinkers, the theory is insensitive to the chemical nature of crosslinkers. This suggests that adding passive crosslinkers in a gel, while keeping high ATP and pluronic concentrations, should modify the instability mode of the gel.  

\subsection{The bending/buckling and local/global contraction transitions are controlled by the concentration of passive linkers}

\begin{figure}
	\includegraphics[width=\linewidth]{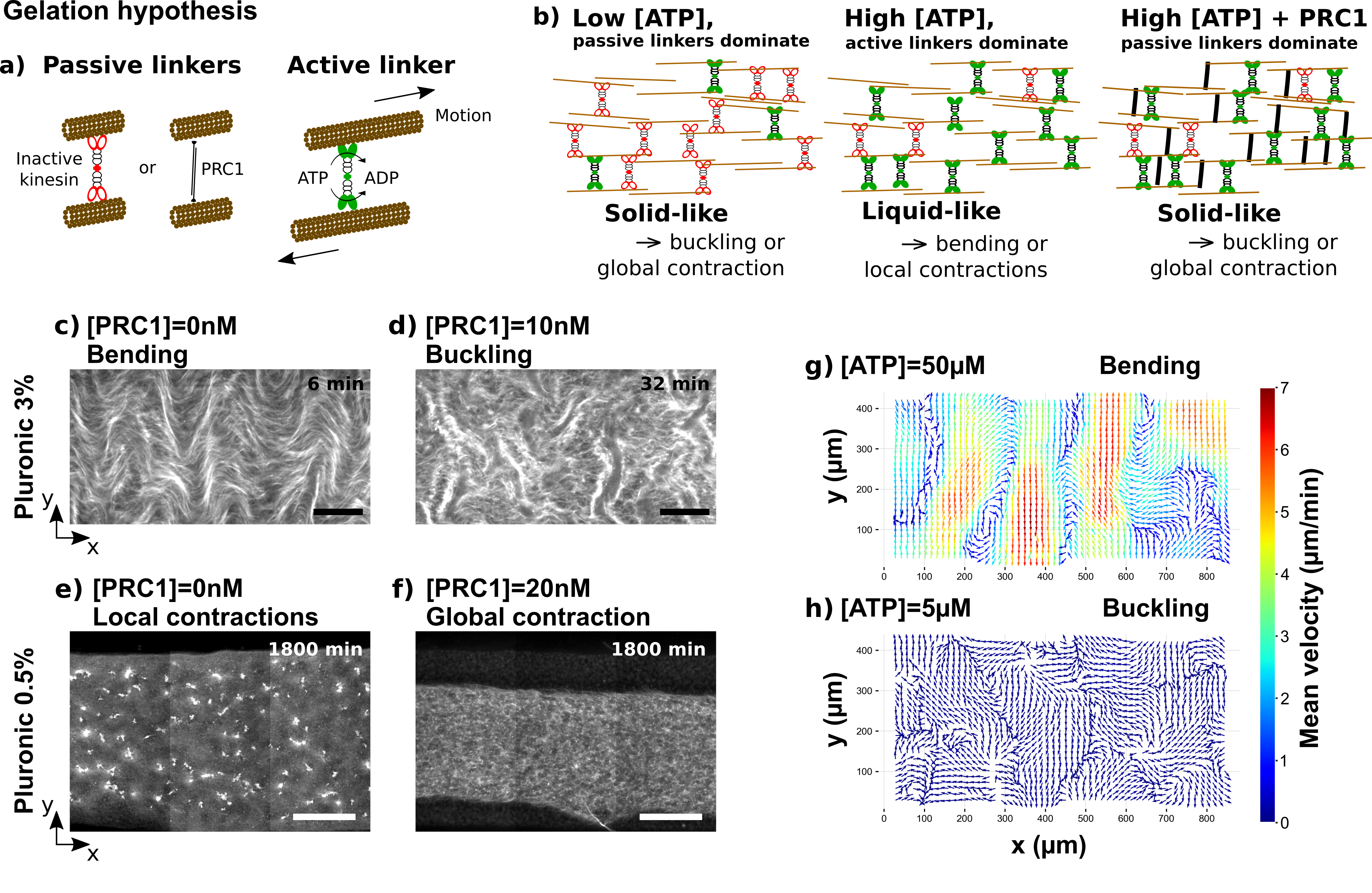}
	\caption{
		The bending/buckling and local/global contraction transitions are controlled by the concentration of passive linkers that induce a gelation transition.
		(a) Schemes of passive linkers (ATP-depleted kinesin, in red, and PRC1 in black) and active one (ATP-consuming kinesin, in green). (b) Schemes of the active gel at different [ATP] and [PRC1] highlighting their liquid/solid-like state and the instabilities observed. 		Epifluorescence images showing the instabilities observed at high [ATP] (50~$\mu$M) and either high (c,d) or low (e,f) pluronic concentration and low (c,e) or high (d,f)  [PRC1].
		Scale bars are 200~$\mu$m for (c,d) and 500~$\mu$m for (e,f). g,h) Particle image velocimetry of the bending (g) and buckling (h) instabilities observed at 0~nM PRC1, 5.5~\% pluronic and [ATP] as indicated. In panels c-f $[\textrm{MT}] = 1$~mg/mL.}
	\label{fig:PRC1}
\end{figure}

To experimentally challenge this last hypothesis, we introduced passive PRC1 linkers\cite{chandrakar_confinement_2020} at constant [ATP]. At high pluronic, we observed that 10 nM PRC1 was sufficient to convert a bending into a buckling instability (Fig.~\ref{fig:PRC1}c,d). At low pluronic, a system displaying local contractions in the absence of PRC1 displayed global ones at 20 nM PRC1 (Fig.~\ref{fig:PRC1}e,f). In some occasions, we observed that a bending state became global contractions when PRC1 was added (Fig.~S\ref{FigSI_bending_became_contraction}). Importantly, the characteristic times of the bending/buckling instabilities, on the one hand, and of the local/global contraction instabilities, on the other hand, were similar at constant [ATP] and different PRC1 (Table~S\ref{si_tab_char_time_PRC1}), in agreement with experiments at $[\textrm{ATP}]_c$ (Fig.~S\ref{FigSI_transition_bending_buckling}) and in contrast with what was observed at different [ATP] in the absence of PRC1 (Table~S\ref{si_tab_char_time}). 

The gelation transition occurred at $[\textrm{ATP}]_c\approx10~\mu$M in Fig.~\ref{PhaseSpace}. Using a Michaelis-Menten model, we can estimate the concentration of passive motors as $c_m^p = c_m^0(1-[\textrm{ATP}]/(K_M+[\textrm{ATP}]))$, where $c_m^0$ is the concentration of all motors and $K_M=40-60~\mu$M is the Michaelis-Menten constant of kinesin-1 motor and ATP\cite{schnitzer_1997, hua_1997}. Taking $c_m^0 = 25$~nM for our experiments, one finds $c_m^p\approx 20$~nM, which corresponds to a concentration of passive linkers $c_l^p\approx 5$~nM (Fig.~S\ref{FigSI_passive_motor_linker_scheme}). This result is consistent with the gelation transition observed at 10 and 20 nM PRC1, respectively for bending/buckling and local/global contractions (Fig.~\ref{fig:PRC1}c$-$f).

To further test the gelation hypothesis, we quantified the average velocity flows of the microtubules in the $xy$-plane during the bending and buckling instabilities at different [ATP] in the absence of PRC1 (Fig.~\ref{fig:PRC1}g,h). During bending, strong flows up to 7 $\mu$m/min arise parallel to the $y$ axis as observed previously\cite{martinez2019, chandrakar_confinement_2020}. In contrast, during buckling the flows do not show a clear structure and they are significantly smaller, on the order of 0.4 $\mu$m/min, as expected for a solid. 

These observations clearly show that a large fraction of kinesin motors act as passive crosslinkers at low [ATP] but not at high [ATP], implying that [ATP] controls a transition from a crosslinked gel to a fluid.

\section{Conclusion}

We have observed four important spatial instabilities -- bending, buckling, local and global contractions -- in a single type of cytoskeletal active gel and we have shown that they are determined by two factors: the crosslinking of the gel and the concentration of depletion agent. Our theoretical and experimental results demonstrate that the bending-to-buckling and local-to-global contraction transitions are determined by a gelation transition in the active gel, induced either by motors acting as passive crosslinkers when [ATP] decreases, or by passive linkers at a fixed [ATP]. However, although the depletant concentration is known to promote filament bundling\cite{Needleman2004}, it is not yet clear how this property controls the bending-to-local contraction and buckling-to-global contraction transitions. 
 
Our work generalizes previous results that showed that the local-to-global contraction transition is controlled by the percolation of the gel induced by passive linkers in actin-myosin gels\cite{alvarado_molecular_2013}. Building on previous work \cite{gagnon_shear-induced_2020}, our results underline the dual role of motors both as passive and active linkers in cytoskeletal active gels. In particular, we show, by varying [ATP], that motors not only control the degree of active drive, as is generally assumed in active fluid theories \cite{martinez2019, senoussi_tunable_2019}, but also the mechanical properties of the gel, leading to qualitatively distinct spatial instabilities and gel morphologies. 
In summary, this work provides a unified view of spatial instabilities in cytoskeletal active gels and provides an experimental and theoretical framework to design active materials with controllable out-of-equilibrium properties \cite{Vyborna2021, senoussi_sciadv_2021, ross_controlling_2019, Tayar2021}.

\section{Materials and methods}

\textbf{Active solution:} Active linkers are clusters of streptavidin and biotinylated kinesin-1. The active solution was composed of fluorescent GMPCPP-stabilized microtubules (at 0.5-1 mg/mL), kinesin and streptavidin (both at a concentration of 25~nM), ATP and pluronic (from 1~$\mu$M to 100~$\mu$M and from 1.5\% to 5.5\%, respectively), an ionic buffer, an antioxydant mix and an ATP regeneration system. \newline
\textbf{Imaging:}  The active solution was injected into rectangular cross-section channels ($22\times1.5\times0.13$~mm) made using passivated glass slides separated by parafilm. Imaging was done by epifluorescence or confocal microscopy at room temperature. \newline
\textbf{Analysis:} Images were reconstructed with Fiji and analyzed with Python scripts, using openpiv package for PIV analysis.
\newline
See Supporting Information for a more detailed description.

\section{Supporting information}
\begin{itemize}
	\item Experimental and theory details, including Figures S1-S7 and Tables S1-S4.
	\item Movie S1, S2 : Time-lapse microscopy of microtubules displaying a bending instability. 
	\item Movie S3, S4 : Time-lapse microscopy of microtubules displaying a buckling instability. 
	\item Movie S5, S6 : Time-lapse epifluorescence microscopy of microtubules displaying local (S5) and global (S6) contractions.	
	\item Movie S7, S8 : Time-lapse epifluorescence microscopy of microtubules displaying different instabilities depending on the concentration of passive linkers at constant [ATP] and pluronic.
\end{itemize}

\section{Acknowledgment}

Z. Gueroui for a kind gift of the K401 plasmid, F. Lam from the microscopy platform at IBPS, L.L. Pontani for providing access to a spinning disk microscope, and A. Senoussi, Y. Vyborna and G. Duclos for insightful discussions. This work has been funded by the European Research Council (ERC) under the European's Union Horizon 2020 programme (grant No 770940, A.E.-T.). A.M. was supported by a TALENT fellowship from CY Cergy Paris Universit\'e. For PRC1 synthesis, we acknowledge support from Brandeis NSF MRSEC, Bioinspired Soft Materials, DMR-2011846. In the concluding stages of our work, we became aware of a complementary, independent effort by the group of Guillaume Duclos who studied instabilities in kinesin-microtubule active fluids \cite{najma_dual_2021}.

\newpage
\providecommand{\latin}[1]{#1}
\makeatletter
\providecommand{\doi}
  {\begingroup\let\do\@makeother\dospecials
  \catcode`\{=1 \catcode`\}=2\doi@aux}
\providecommand{\doi@aux}[1]{\endgroup\texttt{#1}}
\makeatother
\providecommand*\mcitethebibliography{\thebibliography}
\csname @ifundefined\endcsname{endmcitethebibliography}
  {\let\endmcitethebibliography\endthebibliography}{}

\section{Table of contents figure}

\begin{figure}
	\includegraphics{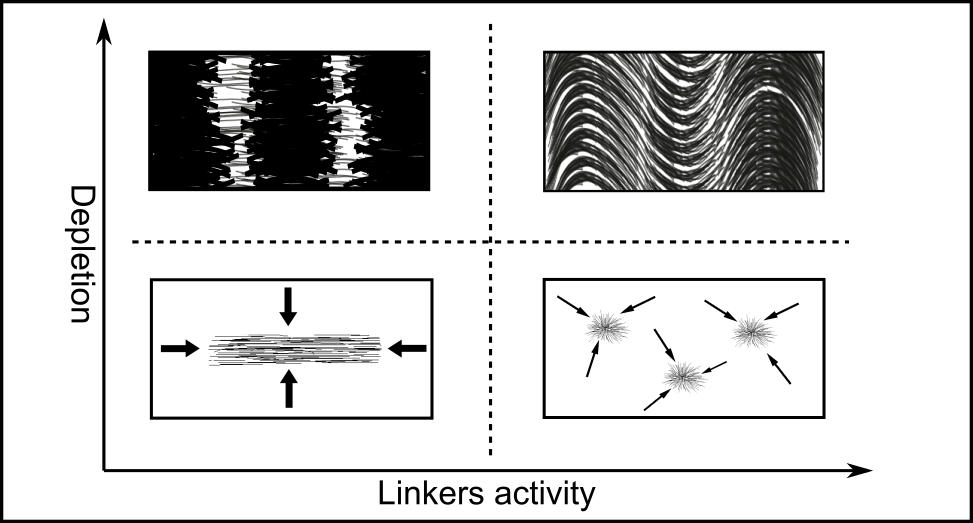}
	\caption{For Table of Contents Only.}
	\label{For Table of Contents Only}
\end{figure}


\begin{mcitethebibliography}{38}
\providecommand*\natexlab[1]{#1}
\providecommand*\mciteSetBstSublistMode[1]{}
\providecommand*\mciteSetBstMaxWidthForm[2]{}
\providecommand*\mciteBstWouldAddEndPuncttrue
  {\def\EndOfBibitem{\unskip.}}
\providecommand*\mciteBstWouldAddEndPunctfalse
  {\let\EndOfBibitem\relax}
\providecommand*\mciteSetBstMidEndSepPunct[3]{}
\providecommand*\mciteSetBstSublistLabelBeginEnd[3]{}
\providecommand*\EndOfBibitem{}
\mciteSetBstSublistMode{f}
\mciteSetBstMaxWidthForm{subitem}{(\alph{mcitesubitemcount})}
\mciteSetBstSublistLabelBeginEnd
  {\mcitemaxwidthsubitemform\space}
  {\relax}
  {\relax}

\bibitem[Marchetti \latin{et~al.}(2013)Marchetti, Joanny, Ramaswamy, Liverpool,
  Prost, Rao, and Simha]{marchetti2013hydrodynamics}
Marchetti,~M.~C.; Joanny,~J.-F.; Ramaswamy,~S.; Liverpool,~T.~B.; Prost,~J.;
  Rao,~M.; Simha,~R.~A. Hydrodynamics of soft active matter. \emph{Reviews of
  Modern Physics} \textbf{2013}, \emph{85}, 1143\relax
\mciteBstWouldAddEndPuncttrue
\mciteSetBstMidEndSepPunct{\mcitedefaultmidpunct}
{\mcitedefaultendpunct}{\mcitedefaultseppunct}\relax
\EndOfBibitem
\bibitem[Needleman and Dogic(2017)Needleman, and Dogic]{needleman_active_2017}
Needleman,~D.; Dogic,~Z. Active matter at the interface between materials
  science and cell biology. \emph{Nature Reviews Materials} \textbf{2017},
  \emph{2}, 17048\relax
\mciteBstWouldAddEndPuncttrue
\mciteSetBstMidEndSepPunct{\mcitedefaultmidpunct}
{\mcitedefaultendpunct}{\mcitedefaultseppunct}\relax
\EndOfBibitem
\bibitem[Bricard \latin{et~al.}(2013)Bricard, Caussin, Desreumaux, Dauchot, and
  Bartolo]{bricard2013}
Bricard,~A.; Caussin,~J.-B.; Desreumaux,~N.; Dauchot,~O.; Bartolo,~D. Emergence
  of macroscopic directed motion in populations of motile colloids.
  \emph{Nature} \textbf{2013}, \emph{503}, 95\relax
\mciteBstWouldAddEndPuncttrue
\mciteSetBstMidEndSepPunct{\mcitedefaultmidpunct}
{\mcitedefaultendpunct}{\mcitedefaultseppunct}\relax
\EndOfBibitem
\bibitem[Chandrakar \latin{et~al.}(2020)Chandrakar, Varghese, Aghvami,
  Baskaran, Dogic, and Duclos]{chandrakar_confinement_2020}
Chandrakar,~P.; Varghese,~M.; Aghvami,~S.; Baskaran,~A.; Dogic,~Z.; Duclos,~G.
  Confinement {Controls} the {Bend} {Instability} of {Three}-{Dimensional}
  {Active} {Liquid} {Crystals}. \emph{Physical Review Letters} \textbf{2020},
  \emph{125}, 257801\relax
\mciteBstWouldAddEndPuncttrue
\mciteSetBstMidEndSepPunct{\mcitedefaultmidpunct}
{\mcitedefaultendpunct}{\mcitedefaultseppunct}\relax
\EndOfBibitem
\bibitem[Senoussi \latin{et~al.}(2019)Senoussi, Kashida, Voituriez, Galas,
  Maitra, and Estevez-Torres]{senoussi_tunable_2019}
Senoussi,~A.; Kashida,~S.; Voituriez,~R.; Galas,~J.-C.; Maitra,~A.;
  Estevez-Torres,~A. Tunable corrugated patterns in an active nematic sheet.
  \emph{Proceedings of the National Academy of Sciences} \textbf{2019},
  \emph{116}, 22464--22470\relax
\mciteBstWouldAddEndPuncttrue
\mciteSetBstMidEndSepPunct{\mcitedefaultmidpunct}
{\mcitedefaultendpunct}{\mcitedefaultseppunct}\relax
\EndOfBibitem
\bibitem[Strübing \latin{et~al.}(2020)Strübing, Khosravanizadeh, Vilfan,
  Bodenschatz, Golestanian, and Guido]{strubing_wrinkling_2020}
Strübing,~T.; Khosravanizadeh,~A.; Vilfan,~A.; Bodenschatz,~E.;
  Golestanian,~R.; Guido,~I. Wrinkling {Instability} in {3D} {Active}
  {Nematics}. \emph{Nano Letters} \textbf{2020}, acs.nanolett.0c01546\relax
\mciteBstWouldAddEndPuncttrue
\mciteSetBstMidEndSepPunct{\mcitedefaultmidpunct}
{\mcitedefaultendpunct}{\mcitedefaultseppunct}\relax
\EndOfBibitem
\bibitem[Ideses \latin{et~al.}(2018)Ideses, Erukhimovitch, Brand, Jourdain,
  Hernandez, Gabinet, Safran, Kruse, and Bernheim-Groswasser]{ideses2018}
Ideses,~Y.; Erukhimovitch,~V.; Brand,~R.; Jourdain,~D.; Hernandez,~J.~S.;
  Gabinet,~U.; Safran,~S.; Kruse,~K.; Bernheim-Groswasser,~A. Spontaneous
  buckling of contractile poroelastic actomyosin sheets. \emph{Nature
  communications} \textbf{2018}, \emph{9}, 2461\relax
\mciteBstWouldAddEndPuncttrue
\mciteSetBstMidEndSepPunct{\mcitedefaultmidpunct}
{\mcitedefaultendpunct}{\mcitedefaultseppunct}\relax
\EndOfBibitem
\bibitem[Bendix \latin{et~al.}(2008)Bendix, Koenderink, Cuvelier, Dogic,
  Koeleman, Brieher, Field, Mahadevan, and Weitz]{bendix_2008}
Bendix,~P.~M.; Koenderink,~G.~H.; Cuvelier,~D.; Dogic,~Z.; Koeleman,~B.~N.;
  Brieher,~W.~M.; Field,~C.~M.; Mahadevan,~L.; Weitz,~D.~A. A quantitative
  analysis of contractility in active cytoskeletal protein networks.
  \emph{Biophysical journal} \textbf{2008}, \emph{94}, 3126--3136\relax
\mciteBstWouldAddEndPuncttrue
\mciteSetBstMidEndSepPunct{\mcitedefaultmidpunct}
{\mcitedefaultendpunct}{\mcitedefaultseppunct}\relax
\EndOfBibitem
\bibitem[Torisawa \latin{et~al.}(2016)Torisawa, Taniguchi, Ishihara, and
  Oiwa]{torisawa_spontaneous_2016}
Torisawa,~T.; Taniguchi,~D.; Ishihara,~S.; Oiwa,~K. Spontaneous {Formation} of
  a {Globally} {Connected} {Contractile} {Network} in a {Microtubule}-{Motor}
  {System}. \emph{Biophysical Journal} \textbf{2016}, \emph{111},
  373--385\relax
\mciteBstWouldAddEndPuncttrue
\mciteSetBstMidEndSepPunct{\mcitedefaultmidpunct}
{\mcitedefaultendpunct}{\mcitedefaultseppunct}\relax
\EndOfBibitem
\bibitem[Alvarado \latin{et~al.}(2013)Alvarado, Sheinman, Sharma, MacKintosh,
  and Koenderink]{alvarado_molecular_2013}
Alvarado,~J.; Sheinman,~M.; Sharma,~A.; MacKintosh,~F.~C.; Koenderink,~G.~H.
  Molecular motors robustly drive active gels to a critically connected state.
  \emph{Nature Physics} \textbf{2013}, \emph{9}, 591--597\relax
\mciteBstWouldAddEndPuncttrue
\mciteSetBstMidEndSepPunct{\mcitedefaultmidpunct}
{\mcitedefaultendpunct}{\mcitedefaultseppunct}\relax
\EndOfBibitem
\bibitem[Foster \latin{et~al.}(2015)Foster, Fuerthauer, Shelley, and
  Needleman]{foster_active_2015}
Foster,~P.~J.; Fuerthauer,~S.; Shelley,~M.~J.; Needleman,~D.~J. Active
  contraction of microtubule networks. \emph{Elife} \textbf{2015}, \emph{4},
  e10837\relax
\mciteBstWouldAddEndPuncttrue
\mciteSetBstMidEndSepPunct{\mcitedefaultmidpunct}
{\mcitedefaultendpunct}{\mcitedefaultseppunct}\relax
\EndOfBibitem
\bibitem[Nédélec \latin{et~al.}(1997)Nédélec, Surrey, Maggs, and
  Leibler]{nedelec_self-organization_1997}
Nédélec,~F.~J.; Surrey,~T.; Maggs,~A.~C.; Leibler,~S. Self-organization of
  microtubules and motors. \emph{Nature} \textbf{1997}, \emph{389},
  305--308\relax
\mciteBstWouldAddEndPuncttrue
\mciteSetBstMidEndSepPunct{\mcitedefaultmidpunct}
{\mcitedefaultendpunct}{\mcitedefaultseppunct}\relax
\EndOfBibitem
\bibitem[Sanchez \latin{et~al.}(2012)Sanchez, Chen, DeCamp, Heymann, and
  Dogic]{sanchez_spontaneous_2012}
Sanchez,~T.; Chen,~D. T.~N.; DeCamp,~S.~J.; Heymann,~M.; Dogic,~Z. Spontaneous
  motion in hierarchically assembled active matter. \emph{Nature}
  \textbf{2012}, \emph{491}, 431--434\relax
\mciteBstWouldAddEndPuncttrue
\mciteSetBstMidEndSepPunct{\mcitedefaultmidpunct}
{\mcitedefaultendpunct}{\mcitedefaultseppunct}\relax
\EndOfBibitem
\bibitem[Mart{\'\i}nez-Prat \latin{et~al.}(2019)Mart{\'\i}nez-Prat,
  Ign{\'e}s-Mullol, Casademunt, and Sagu{\'e}s]{martinez2019}
Mart{\'\i}nez-Prat,~B.; Ign{\'e}s-Mullol,~J.; Casademunt,~J.; Sagu{\'e}s,~F.
  Selection mechanism at the onset of active turbulence. \emph{Nature physics}
  \textbf{2019}, \emph{15}, 362\relax
\mciteBstWouldAddEndPuncttrue
\mciteSetBstMidEndSepPunct{\mcitedefaultmidpunct}
{\mcitedefaultendpunct}{\mcitedefaultseppunct}\relax
\EndOfBibitem
\bibitem[Kumar \latin{et~al.}(2018)Kumar, Zhang, de~Pablo, and
  Gardel]{kumar2018}
Kumar,~N.; Zhang,~R.; de~Pablo,~J.~J.; Gardel,~M.~L. Tunable structure and
  dynamics of active liquid crystals. \emph{Science advances} \textbf{2018},
  \emph{4}, eaat7779\relax
\mciteBstWouldAddEndPuncttrue
\mciteSetBstMidEndSepPunct{\mcitedefaultmidpunct}
{\mcitedefaultendpunct}{\mcitedefaultseppunct}\relax
\EndOfBibitem
\bibitem[Surrey \latin{et~al.}(2001)Surrey, Nedelec, Leibler, and
  Karsenti]{surrey_physical_2001}
Surrey,~T.; Nedelec,~F.; Leibler,~S.; Karsenti,~E. Physical properties
  determining self-organization of motors and microtubules. \emph{Science}
  \textbf{2001}, \emph{292}, 1167--1171\relax
\mciteBstWouldAddEndPuncttrue
\mciteSetBstMidEndSepPunct{\mcitedefaultmidpunct}
{\mcitedefaultendpunct}{\mcitedefaultseppunct}\relax
\EndOfBibitem
\bibitem[Nasirimarekani \latin{et~al.}(2021)Nasirimarekani, Strübing, Vilfan,
  and Guido]{nasirimarekani_tuning_2021}
Nasirimarekani,~V.; Strübing,~T.; Vilfan,~A.; Guido,~I. Tuning the
  {Properties} of {Active} {Microtubule} {Networks} by {Depletion} {Forces}.
  \emph{Langmuir} \textbf{2021}, \emph{37}, 7919--7927\relax
\mciteBstWouldAddEndPuncttrue
\mciteSetBstMidEndSepPunct{\mcitedefaultmidpunct}
{\mcitedefaultendpunct}{\mcitedefaultseppunct}\relax
\EndOfBibitem
\bibitem[Senoussi \latin{et~al.}(2021)Senoussi, Galas, and
  Estevez-Torres]{senoussi_biorxiv_2021}
Senoussi,~A.; Galas,~J.-C.; Estevez-Torres,~A. Programmed mechano-chemical
  coupling in reaction-diffusion active matter. \emph{bioRxiv} \textbf{2021},
  2021.03.13.435232\relax
\mciteBstWouldAddEndPuncttrue
\mciteSetBstMidEndSepPunct{\mcitedefaultmidpunct}
{\mcitedefaultendpunct}{\mcitedefaultseppunct}\relax
\EndOfBibitem
\bibitem[Senoussi \latin{et~al.}(2021)Senoussi, Galas, and
  Estevez-Torres]{senoussi_sciadv_2021}
Senoussi,~A.; Galas,~J.-C.; Estevez-Torres,~A. Programmed mechano-chemical
  coupling in reaction-diffusion active matter. \emph{Science Advances}
  \textbf{2021}, \emph{7}, eabi9865\relax
\mciteBstWouldAddEndPuncttrue
\mciteSetBstMidEndSepPunct{\mcitedefaultmidpunct}
{\mcitedefaultendpunct}{\mcitedefaultseppunct}\relax
\EndOfBibitem
\bibitem[Lemma \latin{et~al.}(2021)Lemma, Mitchell, Subramanian, Needleman, and
  Dogic]{lemma_active_2021}
Lemma,~B.; Mitchell,~N.~P.; Subramanian,~R.; Needleman,~D.~J.; Dogic,~Z. Active
  microphase separation in mixtures of microtubules and tip-accumulating
  molecular motors. \emph{arXiv:2107.12281 [cond-mat, physics:physics]}
  \textbf{2021}, \relax
\mciteBstWouldAddEndPunctfalse
\mciteSetBstMidEndSepPunct{\mcitedefaultmidpunct}
{}{\mcitedefaultseppunct}\relax
\EndOfBibitem
\bibitem[Roostalu \latin{et~al.}(2018)Roostalu, Rickman, Thomas, Nedelec, and
  Surrey]{roostalu_determinants_2018}
Roostalu,~J.; Rickman,~J.; Thomas,~C.; Nedelec,~F.; Surrey,~T. Determinants of
  {Polar} versus {Nematic} {Organization} in {Networks} of {Dynamic}
  {Microtubules} and {Mitotic} {Motors}. \emph{Cell} \textbf{2018}, \emph{175},
  796--+\relax
\mciteBstWouldAddEndPuncttrue
\mciteSetBstMidEndSepPunct{\mcitedefaultmidpunct}
{\mcitedefaultendpunct}{\mcitedefaultseppunct}\relax
\EndOfBibitem
\bibitem[Gagnon \latin{et~al.}(2020)Gagnon, Dessi, Berezney, Boros, Chen,
  Dogic, and Blair]{gagnon_shear-induced_2020}
Gagnon,~D.~A.; Dessi,~C.; Berezney,~J.~P.; Boros,~R.; Chen,~D. T.-N.;
  Dogic,~Z.; Blair,~D.~L. Shear-Induced Gelation of Self-Yielding Active
  Networks. \emph{Phys. Rev. Lett.} \textbf{2020}, \emph{125}, 178003\relax
\mciteBstWouldAddEndPuncttrue
\mciteSetBstMidEndSepPunct{\mcitedefaultmidpunct}
{\mcitedefaultendpunct}{\mcitedefaultseppunct}\relax
\EndOfBibitem
\bibitem[Marchetti \latin{et~al.}(2013)Marchetti, Joanny, Ramaswamy, Liverpool,
  Prost, Rao, and Simha]{RMP}
Marchetti,~M.~C.; Joanny,~J.~F.; Ramaswamy,~S.; Liverpool,~T.~B.; Prost,~J.;
  Rao,~M.; Simha,~R.~A. Hydrodynamics of soft active matter. \emph{Rev. Mod.
  Phys.} \textbf{2013}, \emph{85}, 1143--1189\relax
\mciteBstWouldAddEndPuncttrue
\mciteSetBstMidEndSepPunct{\mcitedefaultmidpunct}
{\mcitedefaultendpunct}{\mcitedefaultseppunct}\relax
\EndOfBibitem
\bibitem[Ramaswamy(2010)]{SR_rev}
Ramaswamy,~S. The Mechanics and Statistics of Active Matter. \emph{Annual
  Review of Condensed Matter Physics} \textbf{2010}, \emph{1}, 323--345\relax
\mciteBstWouldAddEndPuncttrue
\mciteSetBstMidEndSepPunct{\mcitedefaultmidpunct}
{\mcitedefaultendpunct}{\mcitedefaultseppunct}\relax
\EndOfBibitem
\bibitem[Prost \latin{et~al.}(2015)Prost, J{\"u}licher, and Joanny]{Prost2015}
Prost,~J.; J{\"u}licher,~F.; Joanny,~J.-F. Active gel physics. \emph{Nature
  Physics} \textbf{2015}, \emph{11}, 111--117\relax
\mciteBstWouldAddEndPuncttrue
\mciteSetBstMidEndSepPunct{\mcitedefaultmidpunct}
{\mcitedefaultendpunct}{\mcitedefaultseppunct}\relax
\EndOfBibitem
\bibitem[Jülicher \latin{et~al.}(2018)Jülicher, Grill, and
  Salbreux]{SalJul_RepPhys}
Jülicher,~F.; Grill,~S.~W.; Salbreux,~G. Hydrodynamic theory of active matter.
  \emph{Reports on Progress in Physics} \textbf{2018}, \emph{81}, 076601\relax
\mciteBstWouldAddEndPuncttrue
\mciteSetBstMidEndSepPunct{\mcitedefaultmidpunct}
{\mcitedefaultendpunct}{\mcitedefaultseppunct}\relax
\EndOfBibitem
\bibitem[Hemingway \latin{et~al.}(2015)Hemingway, Maitra, Banerjee, Marchetti,
  Ramaswamy, Fielding, and Cates]{Ano_poly}
Hemingway,~E.~J.; Maitra,~A.; Banerjee,~S.; Marchetti,~M.~C.; Ramaswamy,~S.;
  Fielding,~S.~M.; Cates,~M.~E. Active Viscoelastic Matter: From Bacterial Drag
  Reduction to Turbulent Solids. \emph{Phys. Rev. Lett.} \textbf{2015},
  \emph{114}, 098302\relax
\mciteBstWouldAddEndPuncttrue
\mciteSetBstMidEndSepPunct{\mcitedefaultmidpunct}
{\mcitedefaultendpunct}{\mcitedefaultseppunct}\relax
\EndOfBibitem
\bibitem[Beris and Edwards(1994)Beris, and Edwards]{Beris1994}
Beris,~A.~N.; Edwards,~B.~J. \emph{Thermodynamics of Flowing Systems: with
  Internal Microstructure}; Oxford Engineering Science Series; Oxford
  University Press: New York, 1994; p 704\relax
\mciteBstWouldAddEndPuncttrue
\mciteSetBstMidEndSepPunct{\mcitedefaultmidpunct}
{\mcitedefaultendpunct}{\mcitedefaultseppunct}\relax
\EndOfBibitem
\bibitem[Voituriez \latin{et~al.}(2005)Voituriez, Joanny, and Prost]{Voit}
Voituriez,~R.; Joanny,~J.~F.; Prost,~J. Spontaneous flow transition in active
  polar gels. \emph{Europhysics Letters ({EPL})} \textbf{2005}, \emph{70},
  404--410\relax
\mciteBstWouldAddEndPuncttrue
\mciteSetBstMidEndSepPunct{\mcitedefaultmidpunct}
{\mcitedefaultendpunct}{\mcitedefaultseppunct}\relax
\EndOfBibitem
\bibitem[Aditi~Simha and Ramaswamy(2002)Aditi~Simha, and Ramaswamy]{Aditi1}
Aditi~Simha,~R.; Ramaswamy,~S. Hydrodynamic Fluctuations and Instabilities in
  Ordered Suspensions of Self-Propelled Particles. \emph{Phys. Rev. Lett.}
  \textbf{2002}, \emph{89}, 058101\relax
\mciteBstWouldAddEndPuncttrue
\mciteSetBstMidEndSepPunct{\mcitedefaultmidpunct}
{\mcitedefaultendpunct}{\mcitedefaultseppunct}\relax
\EndOfBibitem
\bibitem[Schnitzer and Block(1997)Schnitzer, and Block]{schnitzer_1997}
Schnitzer,~M.~J.; Block,~S.~M. Kinesin hydrolyses one ATP per 8-nm step.
  \emph{Nature} \textbf{1997}, \emph{388}, 386--390\relax
\mciteBstWouldAddEndPuncttrue
\mciteSetBstMidEndSepPunct{\mcitedefaultmidpunct}
{\mcitedefaultendpunct}{\mcitedefaultseppunct}\relax
\EndOfBibitem
\bibitem[Hua \latin{et~al.}(1997)Hua, Young, Fleming, and Gelles]{hua_1997}
Hua,~W.; Young,~E.~C.; Fleming,~M.~L.; Gelles,~J. Coupling of kinesin steps to
  ATP hydrolysis. \emph{Nature} \textbf{1997}, \emph{388}, 390--393\relax
\mciteBstWouldAddEndPuncttrue
\mciteSetBstMidEndSepPunct{\mcitedefaultmidpunct}
{\mcitedefaultendpunct}{\mcitedefaultseppunct}\relax
\EndOfBibitem
\bibitem[Needleman \latin{et~al.}(2004)Needleman, Ojeda-Lopez, Raviv, Ewert,
  Jones, Miller, Wilson, and Safinya]{Needleman2004}
Needleman,~D.~J.; Ojeda-Lopez,~M.~A.; Raviv,~U.; Ewert,~K.; Jones,~J.~B.;
  Miller,~H.~P.; Wilson,~L.; Safinya,~C.~R. Synchrotron X-ray Diffraction Study
  of Microtubules Buckling and Bundling under Osmotic Stress: A Probe of
  Interprotofilament Interactions. \emph{Physical Review Letters}
  \textbf{2004}, \emph{93}, 198104\relax
\mciteBstWouldAddEndPuncttrue
\mciteSetBstMidEndSepPunct{\mcitedefaultmidpunct}
{\mcitedefaultendpunct}{\mcitedefaultseppunct}\relax
\EndOfBibitem
\bibitem[Vyborna \latin{et~al.}(2021)Vyborna, Galas, and
  Estevez-Torres]{Vyborna2021}
Vyborna,~Y.; Galas,~J.-C.; Estevez-Torres,~A. DNA-Controlled Spatiotemporal
  Patterning of a Cytoskeletal Active Gel. \emph{Journal of the American
  Chemical Society} \textbf{2021}, \emph{143}, 20022–20026\relax
\mciteBstWouldAddEndPuncttrue
\mciteSetBstMidEndSepPunct{\mcitedefaultmidpunct}
{\mcitedefaultendpunct}{\mcitedefaultseppunct}\relax
\EndOfBibitem
\bibitem[Ross \latin{et~al.}(2019)Ross, Lee, Qu, Banks, Phillips, and
  Thomson]{ross_controlling_2019}
Ross,~T.~D.; Lee,~H.~J.; Qu,~Z.; Banks,~R.~A.; Phillips,~R.; Thomson,~M.
  Controlling {Organization} and {Forces} in {Active} {Matter} {Through}
  {Optically}-{Defined} {Boundaries}. \emph{Nature} \textbf{2019}, \emph{572},
  224--229, arXiv: 1812.09418\relax
\mciteBstWouldAddEndPuncttrue
\mciteSetBstMidEndSepPunct{\mcitedefaultmidpunct}
{\mcitedefaultendpunct}{\mcitedefaultseppunct}\relax
\EndOfBibitem
\bibitem[Tayar \latin{et~al.}(2021)Tayar, Hagan, and Dogic]{Tayar2021}
Tayar,~A.~M.; Hagan,~M.~F.; Dogic,~Z. Active liquid crystals powered by
  force-sensing DNA-motor clusters. \emph{Proceedings of the National Academy
  of Sciences} \textbf{2021}, \emph{118}, e2102873118\relax
\mciteBstWouldAddEndPuncttrue
\mciteSetBstMidEndSepPunct{\mcitedefaultmidpunct}
{\mcitedefaultendpunct}{\mcitedefaultseppunct}\relax
\EndOfBibitem
\bibitem[Najma \latin{et~al.}(2021)Najma, Varghese, Tsidilkovski, Lemma,
  Baskaran, and Duclos]{najma_dual_2021}
Najma,~B.; Varghese,~M.; Tsidilkovski,~L.; Lemma,~L.; Baskaran,~A.; Duclos,~G.
  Dual antagonistic role of motor proteins in fluidizing active networks.
  \emph{arXiv:2112.11364 [cond-mat, physics:physics]} \textbf{2021}, \relax
\mciteBstWouldAddEndPunctfalse
\mciteSetBstMidEndSepPunct{\mcitedefaultmidpunct}
{}{\mcitedefaultseppunct}\relax
\EndOfBibitem
\end{mcitethebibliography}
\end{document}


\newpage
\setcounter{tocdepth}{2}
\tableofcontents
\newpage

\section{Methods\label{SI_methods}}

\subsection{Chemicals and reagents}

All chemicals and reagents were purchased from Sigma-Aldrich, New England Biolabs, Roche, and ThermoScientific.

\subsection{Active and passive linkers}

\textbf{Kinesin K401} - Kinesins were purified as previously described \cite{senoussi_tunable_2019}.
 
\noindent\textbf{Passive crosslinker PRC1} - PRC1 was kindly sent by Guillaume Duclos, Shibani Dalal and Radhika Subramanian. We acknowledge support from Brandeis NSF MRSEC,
241 Bioinspired Soft Materials, DMR-2011846 \cite{subramanian_insights_2010}.

\subsection{Microtubule polymerization}

Tubulin and TRITC-labeled tubulin were purchased from Cytoskeleton, dissolved at 10 mg/mL in 1X PEM buffer (80 mM PIPES pH 6.8, 1 mM EGTA, 1 mM MgSO$_{4}$), flash-frozen and stored at $-80~^{\circ}$C. 

\textbf{Taxol-stabilized microtubules} -
The polymerization mix consists of 1X PEM, 1 mM GTP, 10 \% (w/v) glycerol and tubulin at 5 mg/mL (including 2.5 \% fluorescent tubulin). The mix was incubated at 37 $^{\circ}$C for 20 min. 20 $\mu$M of paclitaxel (in the following taxol) was added to the mix and let at 37 $^{\circ}$C for five more minutes. After polymerization, newly formed microtubules were centrifugated at room temperature for 10 min at 12000 g to remove free tubulin monomers. The microtubules were redissolved into 1X PEM, 1 mM GTP, 10 \% glycerol, 20 $\mu$M taxol and kept in the dark at room temperature. They were used within three days.
\textbf{GMPCPP microtubules} -
The polymerization mix consists of 1X PEM, 0.60~mM GMPCPP (Jena Bioscience), 10 \% (w/v) glycerol, 0.2~mM DTT and tubulin at 5 mg/mL (including 2.5 \% fluorescent tubulin). The mix was incubated at 37 $^{\circ}$C for 30 min and left at room temperature for 5.5 hours. Then the microtubules were flash-frozen and kept at -80~$^{\circ}$C.

\subsection{Experimental conditions}

The active mix is composed of : 
\begin{itemize}
	\item an ionic buffer: 1X PEM buffer (80mM~PIPES, 1~mM EGTA, 1~mM MgSO$_{4}$, 130~mM KOH) supplemented with 3~mM MgSO$_{4}$
	\item an antioxydant mix: 1~mM trolox, 20~mM D-glucose, 3~mM DTT, 150~$\mu$g/mL glucose oxidase, 25~$\mu$g/mL catalase, 0.5~mg/mL BSA
	\item an ATP regenerative system: 5~$\mu$g/mL creatine kinase, 0~mM creatine phosphate
	\item motor clusters: 25~nM K401 kinesin, 25~nM streptavidin 
	\item ATP, Pluronic and microtubules, depending on the experiments, as described in the table below. When taxol-stabilized microtubules are used, 20~$\mu$M taxol is added to the mix.
	
\end{itemize}

\begin{table}[!ht]
\renewcommand{\tablename}{Table S\!\!}\caption{Details of the experimental conditions in each figure of the paper.}\label{Tab_SI_exp_cond} 
	\centering
	\begin{tabular}{|p{3cm}|p{2.5cm}|p{4cm}|p{3.5cm}|}
		\hline
		Figures  & [Microtubules] & ATP & Pluronic~(w/v \%) \\ \hline
		Fig 1 and Fig 5g,h & 0,5~mg/mL GMPCPP & 5~$\mu$M (buckling) and 50~$\mu$M (bending) & 5.5 \\ \hline
		Fig 2, 3, S2, S3, S5  & 0,5~mg/mL GMPCPP & 1~$\mu$M to 100$\mu$M & 1.5 to 5.5\\ \hline
		Fig 4 & 0,5~mg/mL GMPCPP & 5~$\mu$M and 50~$\mu$M of caged-ATP & 5.5 \\ \hline
		Fig 5c-f & 1~mg/mL GMPCPP & 50~$\mu$M & 0.5 (e,f) and 3 (c,d) \\ \hline
		Fig S7 & 0.5~mg/mL GMPCPP & 50~$\mu$M & 3 \\ \hline
		Fig S8 & 0,5~mg/mL taxol & 1~$\mu$M to 10~mM & 0.5 to 2 \\ \hline
	\end{tabular}
\end{table}

Note that control experiments performed with 0~$\mu$M ATP still showed activity (albeit low), suggesting that residual ATP remains from kinesin purification buffer.

\subsection{Channel assembly and imaging}

Channels were assembled using a microscope glass slide (26 x 75 x 1 mm) and a coverslip (22 x 50 x 0.17 mm) separated by stripes of Parafilm cut with a Graphtec Cutting Plotter CE6000-40. Both microscope glass slides and coverlips were passivated using an acrylamide brush \cite{sanchez_spontaneous_2012}. The active mix was filled in the flow cell (22 x 1.5 x 0.130 mm) by capillarity and sealed with vacuum grease.

Epifluorescence images were obtained with a Zeiss Observer 7 automated microscope equipped with a Hamamatsu C9100-02 camera, a 10X objective, a motorized stage and controlled with MicroManager 1.4. Images were recorded automatically using an excitation at 550 nm with a CoolLED pE2. 

Confocal images were obtained with a Leica TCS SP5 II confocal microscope with a 25x water-immersion objective or a X-Light V2 Spinning Disk Confocal system mounted on an upright Nikon Eclipse 80i microscope with a 10x objective.

\subsection{PIV analysis}

Microtubule tracking was done with homemade Python scripts using package openpiv with an interrogation window and a search area size of 64 pixels and an overlap of 48 pixels. The results were validated only if they had a signal to noise value superior to 1.05. The mean was then taken over the instability duration, while removing the 5\% of the strongest speed values.

\newpage

\section{Literature survey of spatial instabilities observed experimentally in 3D cytoskeletal active matter}

In the non-exhaustive table below, we summarize the disparity of conditions in which spatial instabilities have been observed in (3D configuration) cytoskeletal active matter:

\begin{table}
	
\scriptsize
	
	\renewcommand{\tablename}{Table S\!\!}\caption{Observed spatial instabilities (Chaotic flow, bending, buckling, global and local contractions). PEG: Poly(ethylene glycol). PRC1: Protein regulator of cytokinesis 1. }\label{Tab_SI_biblio} 
	\begin{center}
		\begin{tabular}{|p{2cm}|p{3cm}|p{2.8cm}|p{1.5cm}|p{1.5cm}|p{3.3cm}|}
			\hline
			
			\textbf{Reference} &\textbf{Motors}	&	\textbf{Filaments}	&	\textbf{Depletant} & \textbf{Passive crosslinker} & \textbf{Instability} \\
						\hline
						\hline

			Nédélec et al. 1997\cite{nedelec_self-organization_1997}&Kinesin-1/streptavidin clusters 		&Taxol-stabilized microtubules	&  
			&	&Local contractions (asters)	 \\
			\hline		
	
			Surrey et al. 2001\cite{surrey_physical_2001}&Kinesin-1/streptavidin clusters and Ncd		&Taxol-stabilized microtubules	&  
			&	&Local contractions (asters)	 \\
			\hline	
	
			Bendix et al. 2008\cite{bendix_2008} &Myosin II		&Actin	& 
			&$\alpha$-actinin 	&Global contraction 	 \\
			\hline
			
			Sanchez et al. 2012\cite{sanchez_spontaneous_2012} &Kinesin-1/streptavidin clusters   	&GMPCPP-stabilized microtubules	& PEG
			& 			&Chaotic flow \\
			\hline
			
			Alvarado et al. 2013\cite{alvarado_molecular_2013}&Myosin II		&Actin	&  
			&Fascin	&Transition between local and global contractions	 \\
			\hline			
			
			Foster et al. 2015\cite{foster_active_2015}&Xenopus oocyte extract (with dynein and kinesin-5)		&Taxol-stabilized microtubules	&  
			&	&Global contraction	 \\
			\hline			
			
			Torisawa et al. 2016\cite{torisawa_spontaneous_2016}&Tetrameric kinesin Eg5		&Taxol-stabilized microtubules	&  
			&	&Transition between local and global contractions	 \\
			\hline						
			
			Ideses et al. 2018\cite{ideses2018} &Myosin II	clusters	&Actin	& 
			&	Fascin &Buckling 	 \\
			\hline
			
			Roostalu et al. 2018\cite{roostalu_determinants_2018}&Multiheaded kinesin-5 and kinesin-14 		&Dynamic microtubules	&  
			&	&Transition between local contractions and chaotic flow	 \\
			\hline			
			
			Senoussi et al. 2019\cite{senoussi_tunable_2019}&Kinesin 1 with SNAP tag (forms non-specific clusters) and Kinesin-1/streptavidin clusters		&Taxol-stabilized microtubules 	& Pluronic 
			&	&Buckling before chaotic flow	 \\
			\hline			
			
			Strubing et al. 2020\cite{strubing_wrinkling_2020} &Kinesin-1/streptavidin	clusters	&Taxol-stabilized microtubules	& PEG 
			&	&Global contraction and buckling before chaotic flow	 \\
			\hline			
						
	  	    Chandrakar et al. 2020\cite{chandrakar_confinement_2020} &Kinesin-1/streptavidin clusters  	&GMPCPP-stabilized microtubules	& PEG
		    &PRC1 			&Bending \\
		    \hline			
			
			Lemma et al. 2021\cite{lemma_active_2021} &Dimeric kinesin-4  	&GMPCPP-stabilized microtubules	& PEG
			& 			&Transition between local and global contractions, bending and active foams \\
			\hline			
			
			Nasirimarekani et al. 2021\cite{nasirimarekani_tuning_2021} &Kinesin-1/streptavidin clusters		&Taxol-stabilized microtubules	& PEG and Pluronic
			&	&Global contraction 	 \\
			\hline				
	
			Senoussi et al. 2021\cite{senoussi_biorxiv_2021}&Kinesin 1 with SNAP tag (forms non-specific clusters) and Kinesin-1/streptavidin clusters		&Taxol-stabilized microtubules  and GMPCPP-stabilized microtubules	& Pluronic 
			&	&Local and global contractions,  buckling and
			chaotic flow	 \\
			\hline

			Current paper &Kinesin-1/streptavidin clusters	&
			GMPCPP-stabilized microtubules	and taxol-stabilized microtubules  & Pluronic 
			&PRC1 and inactive kinesin clusters	&Transition between  local and global contractions, bending, buckling and chaotic flow 
			 \\
			\hline

		\end{tabular}
	\end{center}
\end{table}

\normalsize

\newpage
\section{Dynamics of pattern formation}

Below, we summarized the methods used to extract a characteristic time for the bending, buckling, global contraction and local contraction experiments.

For global contraction, we measured the width of the gel over time, and fit the contraction using a sigmoid function.

For local contractions, we measured the size of the largest microtubule clusters (typically the 5 largest clusters in a field of view) over time, and fit the cluster mean size using a sigmoid function.

For bending and buckling experiments, we calculated the standard deviation of the microtubule fluorescence over time, and fit the data using a sigmoid function.

\begin{figure}[H]
	\includegraphics[width=1\linewidth]{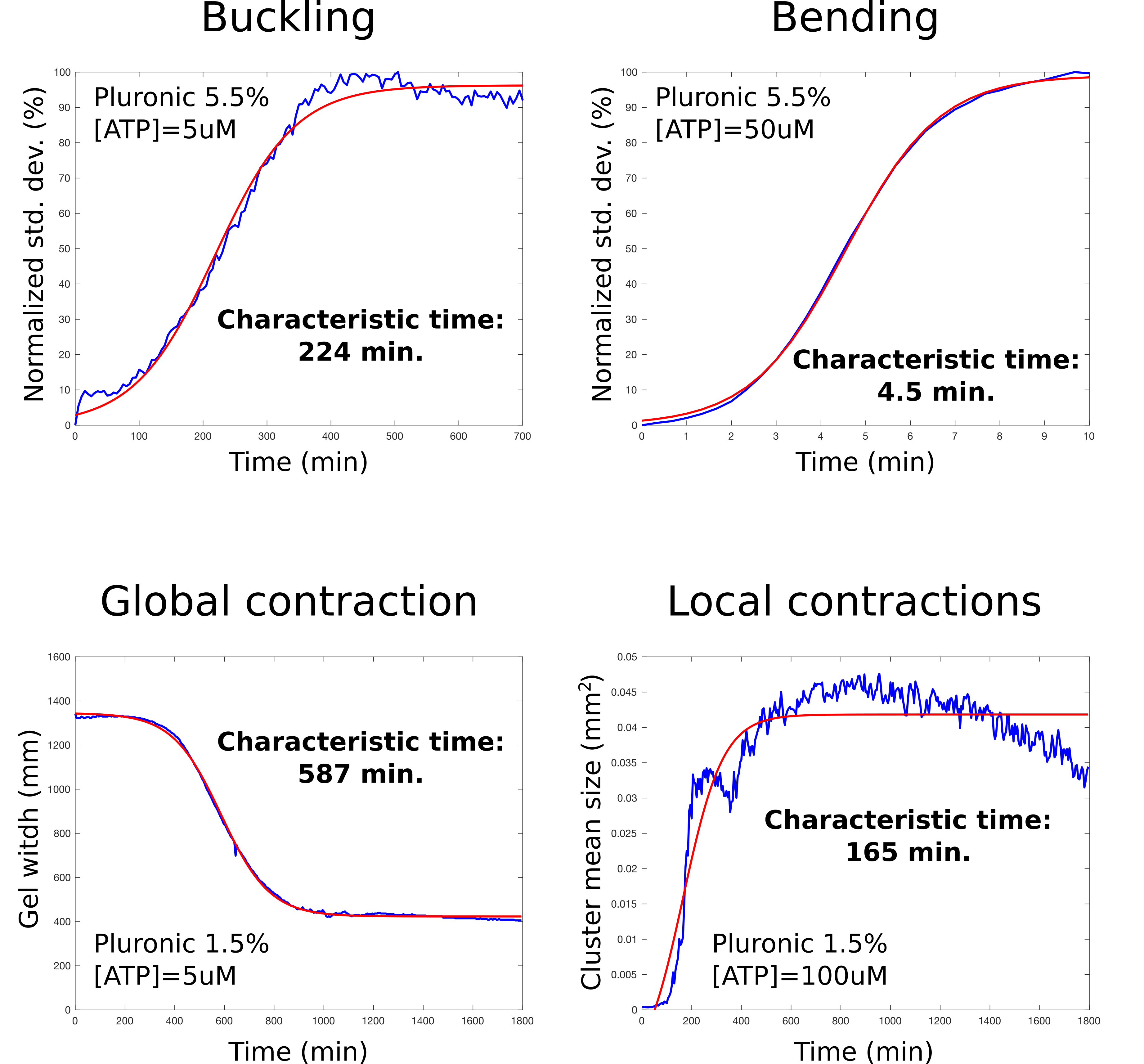}
	\renewcommand{\figurename}{Figure S\!\!}
	\caption{
		Determination of a characteristic time for the bending, buckling, global contraction and local contraction experiments fitting a sigmoid function (in red) to the data (in blue).
	}
	\label{caract_time}
\end{figure}

We provide Fig.~S\ref{caract_time} an example for the four different instabilities.


%
%
%
%
%
%
%
%
%


\maketitle
\begin{figure}[H]
	\includegraphics[width=0.7\textwidth]{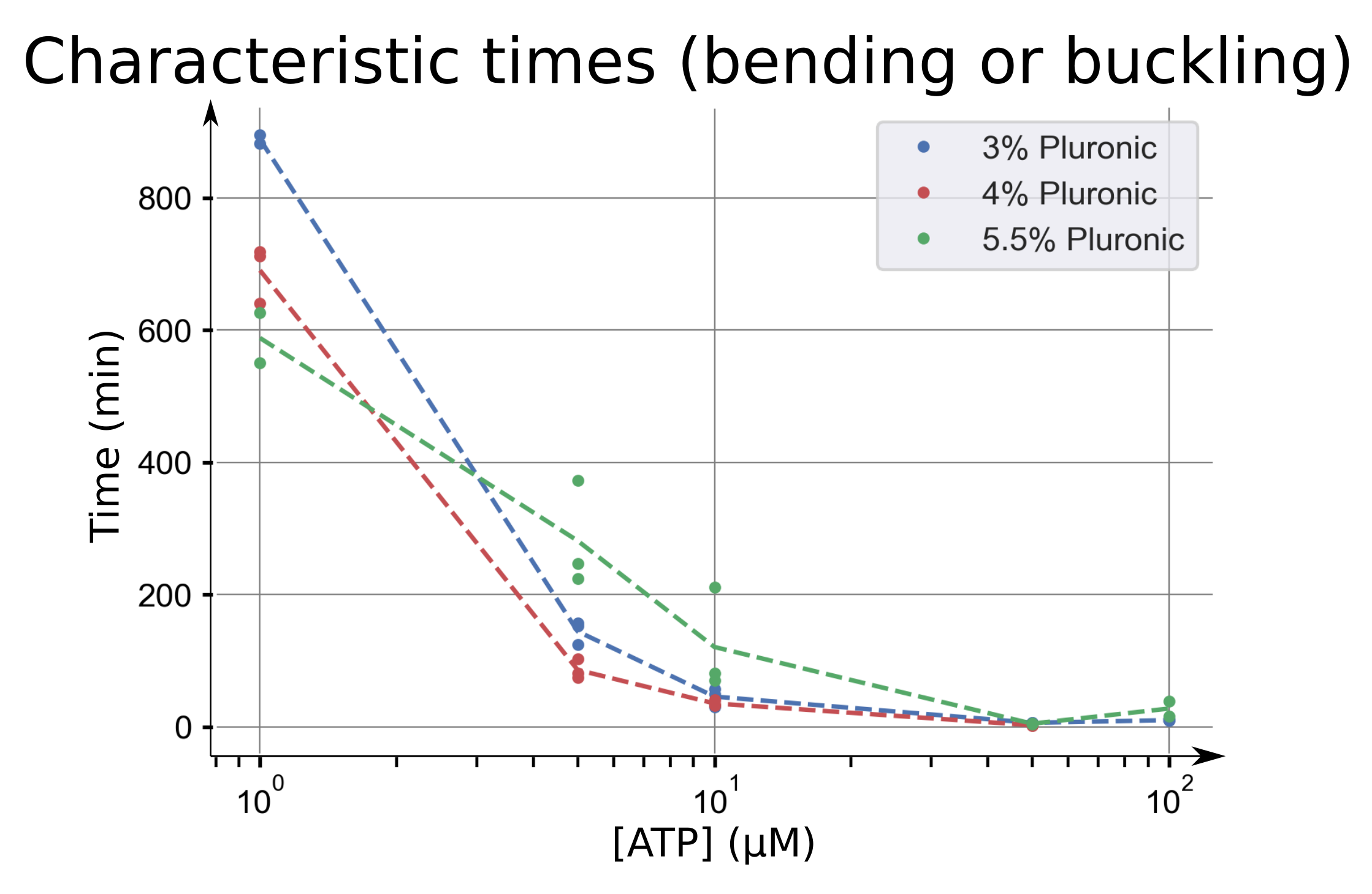}
	\renewcommand{\figurename}{Figure S\!\!}
	\caption{Characteristic times of extensile instabilities (bending and buckling) at different ATP and pluronic concentrations. Time is taken according to the method described above, with triplicates for each pair of conditions (Pluronic and ATP). Dotted lines are a guide to the eye that link the means of these measures.}
	\label{FigSI_BendBuck_overTime}
\end{figure}

\begin{table}
	\renewcommand{\tablename}{Table S\!\!}\caption{Characteristic time for the bending (green), buckling (red), global contraction (yellow) and local contraction (violet) experiments. Associated to Figures 1, 2 and 3 in the main text. [MT]=0.5 mg/mL. } 
	\label{si_tab_char_time}
	\begin{center}
			\begin{tabular}{{c} | {c} | {c}}
			\hline
			&[ATP] (5$\mu$M) 	&[ATP] (50$\mu$M)  \\
			\hline
			Pluronic ($\%$, w/v)&Time (min) &Time (min) \\
			\hline  
					&\cellcolor{red!25}Buckling	  	&\cellcolor{green!25}Bending  	\\			
			3		&\cellcolor{red!25}$144\pm18$	  	&\cellcolor{green!25}$6\pm1$  	\\
			4		&\cellcolor{red!25}$85\pm15$		  	&\cellcolor{green!25}$2\pm1$  	\\
			5.5		&\cellcolor{red!25}$281\pm80$	  	&\cellcolor{green!25}$5\pm1$  	\\ 
			\hline
			&\cellcolor{yellow!25}Global contraction	  	&\cellcolor{blue!20}Local contraction  	\\
			1.5		&\cellcolor{yellow!25}587		  	&\cellcolor{blue!20}156  	\\

			\hline
		\end{tabular}
	\end{center}
\end{table}

\newpage
\section{At [ATP]$_c$, both bending and buckling instabilities are observed, with similar dynamics}

The transition between bending and buckling instabilities happens in the interval 5 $\mu$M - 10 $\mu$M ATP. In this range, replicate experiments yield either one or the other instability or a mixture of the two.

\begin{figure}[H]
	\includegraphics[width=0.8\textwidth]{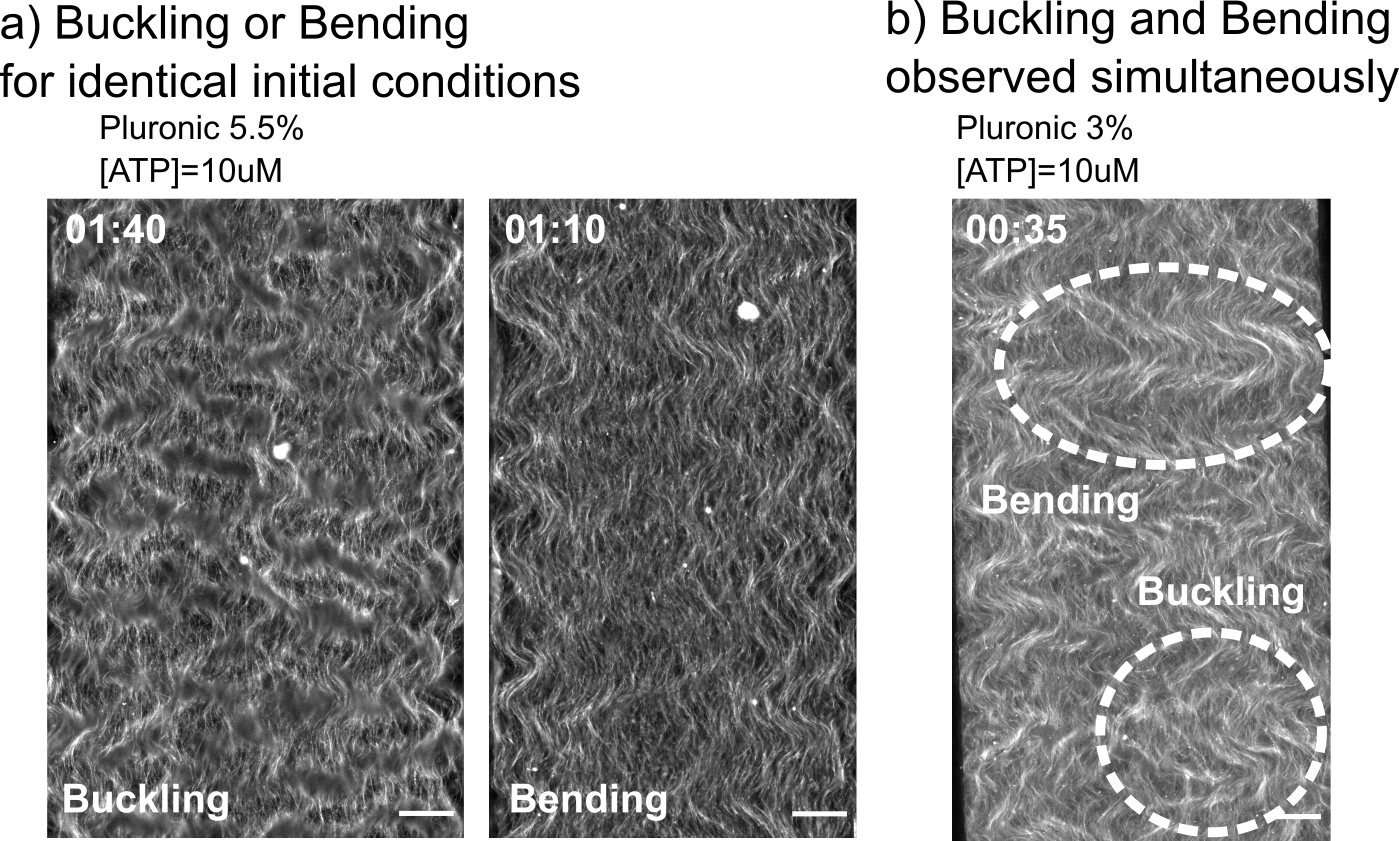}
	\renewcommand{\figurename}{Figure S\!\!}
	\caption{Transition between bending and buckling instabilities. a) Buckling and bending observed for identical initial conditions (Pluronic 5.5\% and [ATP]=10 $\mu$M). b) Buckling and bending observed simultaneously (Pluronic 3\% and [ATP]=10 $\mu$M). Time in h:min. Scale bars are 200 $\mu$m.}
	\label{FigSI_transition_bending_buckling}
\end{figure}

\newpage
\section{Phase space of active fluid instabilities using long taxol-stabilized microtubules}

\begin{figure}[H]
	\includegraphics[width=0.7\textwidth]{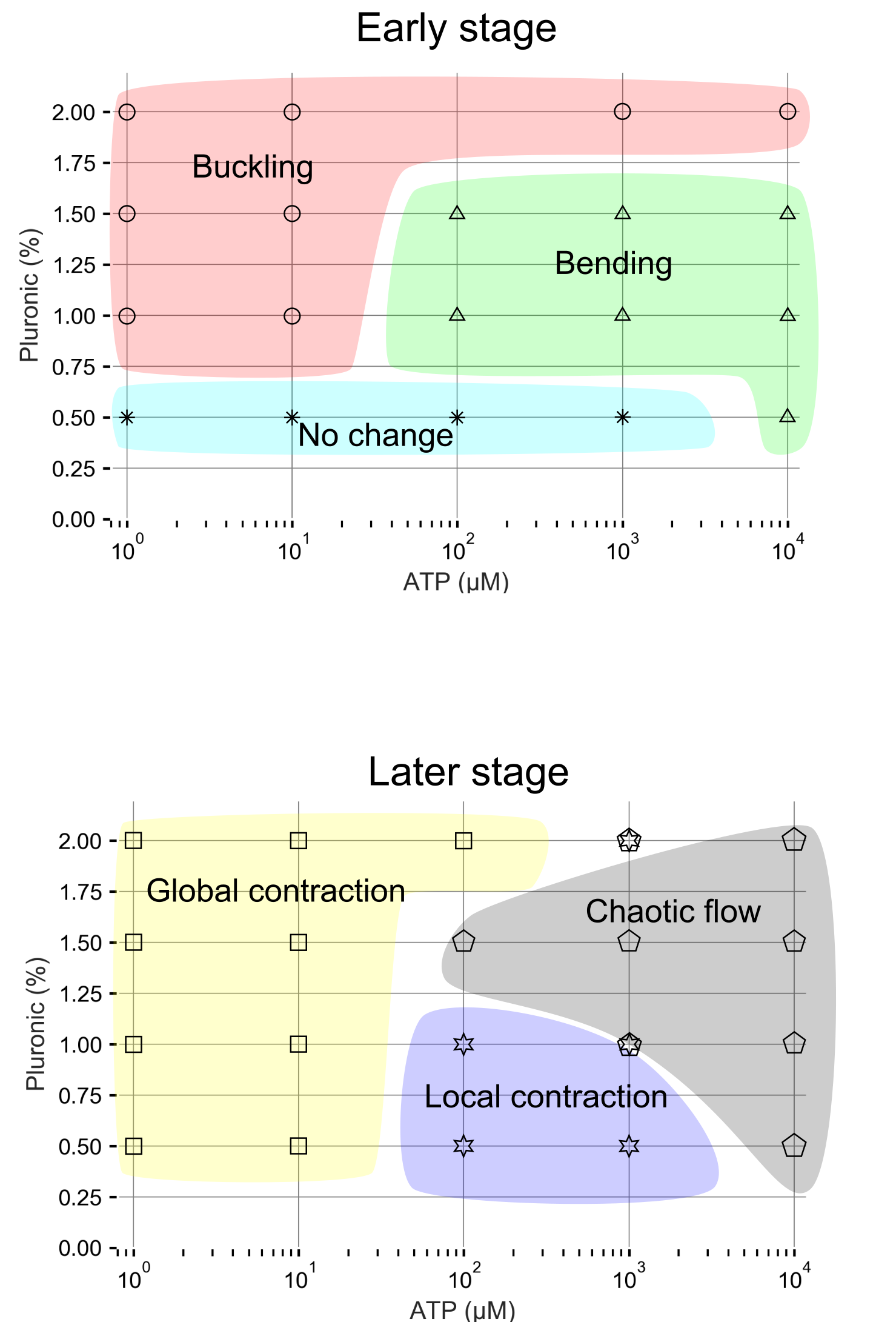}
	\renewcommand{\figurename}{Figure S\!\!}
	\caption{Phase spaces of active fluid instabilities using long taxol-stabilized microtubules as a function of pluronic and ATP
		concentrations, at short and long times. The symbols indicate the observed instabilities and
		the colors are a guide to the eye (The color code is identical as the one used in the main text).}
	\label{FigSI_PhaseSpaceTaxol}
\end{figure}

\newpage
\section{Dynamics of passive contraction}

\begin{figure}[H]
	\includegraphics[width=0.6\textwidth]{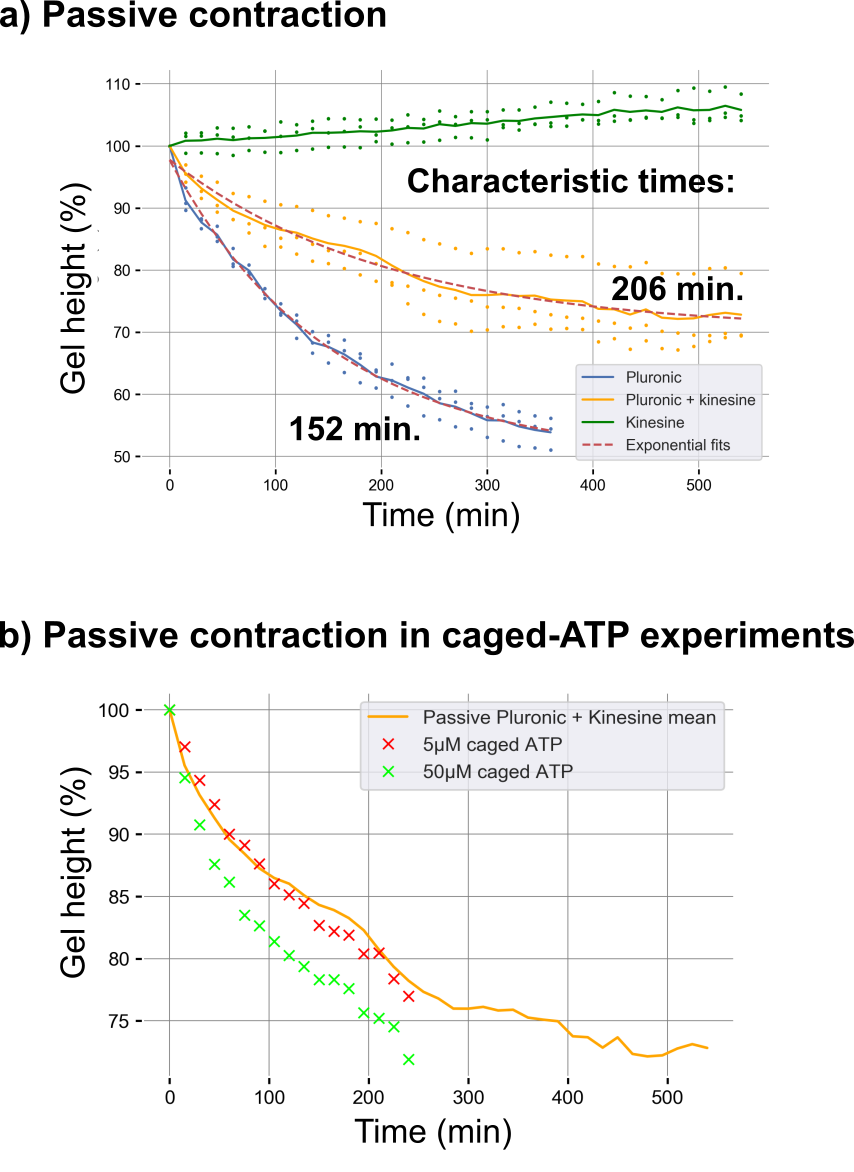}
	\renewcommand{\figurename}{Figure S\!\!}
	\caption{Passive depletion forces induce a passive contraction of the microtubules along z. a) Passive contraction induced by kinesin, pluronic or both. Exponential fits provide characteristic times of 206 minutes for kinesin+pluronic induced contraction, and 152 minutes for pluronic induced contraction. b) Passive contraction measured when caged-ATP is added to the solution. No significant difference in the contraction is observed. \comment{on donne aucune concentration ?}}
	\label{FigSI_PassiveContraction}
\end{figure}

In the absence of ATP, passive depletion forces induce a passive contraction of the microtubules along $z$, whose dynamics are displayed in Fig.~S\ref{FigSI_PassiveContraction}a. To perform bending/buckling experiments while allowing passive contraction to occur at the beginning of the experiment, we used caged-ATP, and UV uncaging to initiate the motor activity. The passive contraction that occurs with caged-ATP is similar to that without ATP (Fig.~S\ref{FigSI_PassiveContraction}b).

\newpage
\section{Experiments with PRC1 passive linkers}

\subsection{The characteristic times of different instabilities remain unchanged at constant [ATP] and [pluronic] as [PRC1] changes}

In Fig.~S\ref{FigSI_BendBuck_overTime} we see that the characteristic time of the instability changes strongly with [ATP]. In particular, in Table~S\ref{si_tab_char_time} we see that bending and local contraction instabilities taking place at 50~$\mu$M ATP are fast compared with their counterparts buckling and global contraction arising at 5~$\mu$M ATP. This could suggest that the former are necessarily faster than the latter. However, the results in Fig.~S\ref{FigSI_transition_bending_buckling} at the critical concentration $[\textrm{ATP}]_c=10$~nM, where bending and buckling are observed simultaneously, indicate that bending  and buckling have similar dynamics at constant [ATP]. In addition, results in Tab.~S\ref{si_tab_char_time_PRC1} at constant [ATP] but different [PRC1] show again that bending and buckling have similar dynamics at constant [ATP] ($4\pm2$ and $14\pm12$ min, respectively). The same is observed for local/global contractions (886 and 830 min, respectively).

\begin{table}
	\renewcommand{\tablename}{Table S\!\!}\caption{Characteristic time for the bending, buckling, global contraction and local contraction experiments with and without the addition of PRC1 passive linkers.} 
	\label{si_tab_char_time_PRC1}
	\begin{center}
		\begin{tabular}{ p{1cm}p{2cm}  p{1cm}l  p{1.5cm}l }
			\hline
			[ATP] ($\mu$M)&Pluronic (\%, w/v)&Time (min) &Pattern&Time (min) &Pattern \\
			\hline
			\hline
			&	&\multicolumn{2}{c}{PRC1 0 nM}	&\multicolumn{2}{c}{PRC1 10 nM} \\
			\cmidrule(lr){3-4}\cmidrule(lr){5-6}
			50	&3.0	&$4\pm2$		&Bending			&$14\pm12$	&Buckling \\
			5	&3.0	&$77\pm5$	&Buckling			&		& \\
			&	&\multicolumn{2}{c}{PRC1 0 nM}	&\multicolumn{2}{c}{PRC1 20 nM} \\
			\cmidrule(lr){3-4}\cmidrule(lr){5-6}
			50	&0.5	&886			&Local cont.		&830		&Global cont. \\
			\hline
		\end{tabular}
	\end{center}
\end{table}

\subsection{Atypical observation when a bending state
	 became a global contraction when PRC1 was added}

\begin{figure}[H]
	\includegraphics[width=0.8\textwidth]{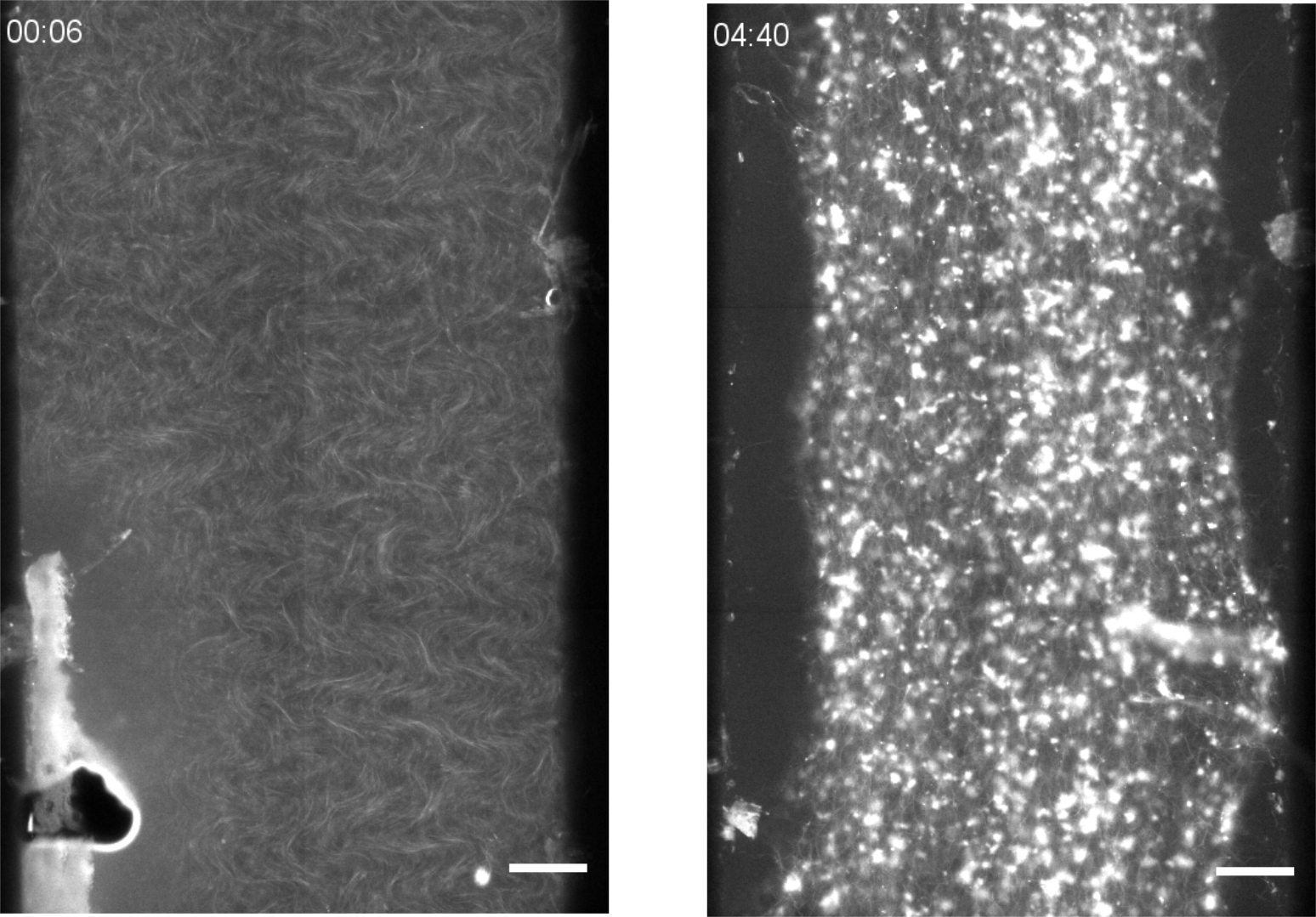}
	\renewcommand{\figurename}{Figure S\!\!}
	\caption{Atypical observation when a bending state
		became a global contraction when PRC1 was added. 3\% Pluronic, [ATP]=50$\mu$M, [PRC1]=0 nM (left) and [PRC1]=10 nM (right). Time in h:min. Scale bars are 200 $\mu$m.}
	\label{FigSI_bending_became_contraction}
\end{figure}

\newpage
\section{Estimating the concentration of passive motor linkers}

\begin{figure}[H]
	\includegraphics[width=0.6\textwidth]{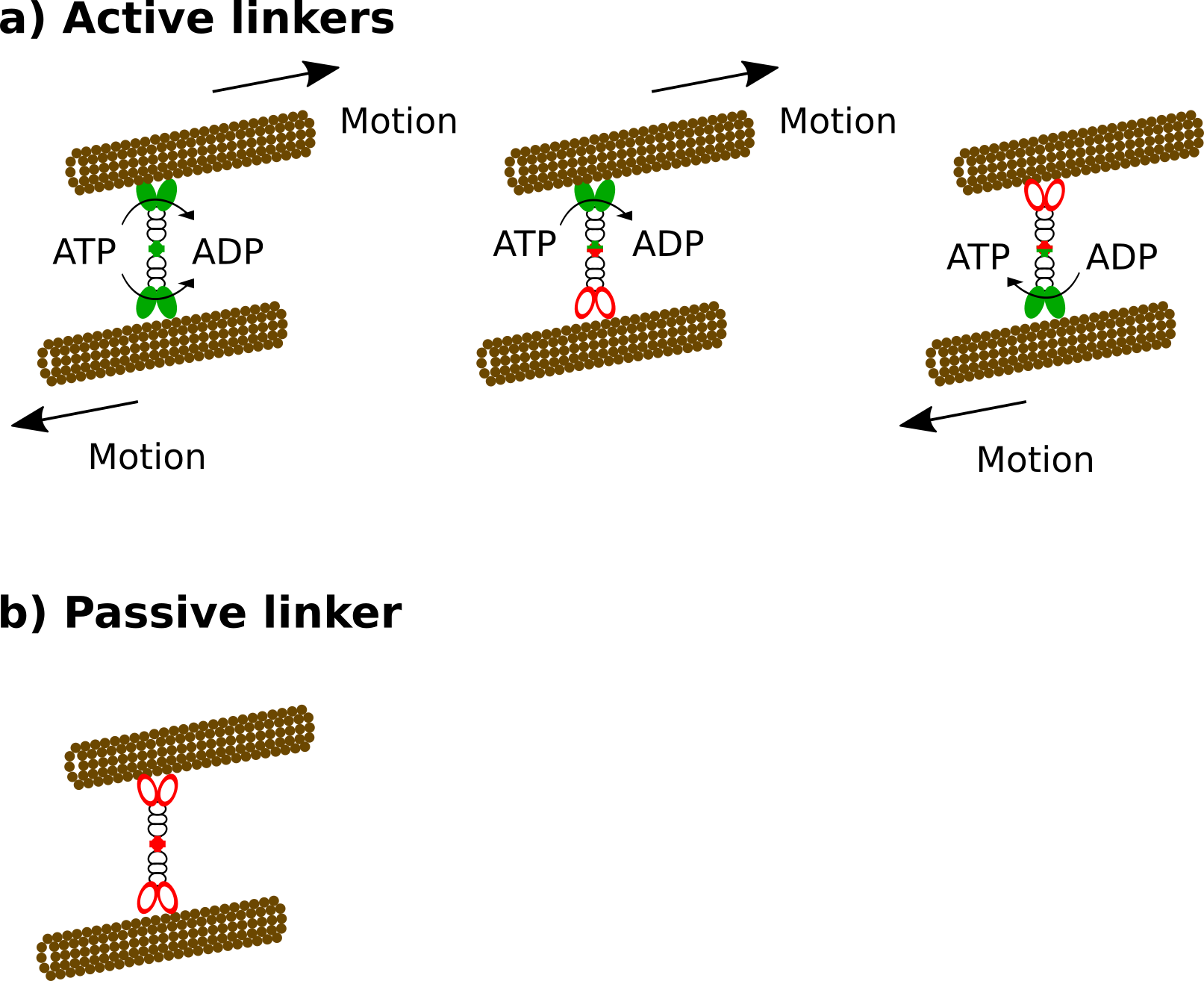}
	\renewcommand{\figurename}{Figure S\!\!}
	\caption{Linkers consisting of aggregated motors are passive if none of the motors in a cluster is active. a) Active linkers may be composed of 2 active motors (left), noted $aa$, or one active and one inactive (center and right), noted $ap$. b) Passive linkers are composed of two passive motors, noted $pp$. With the notations above we have $c_m^a = c_m^{aa} + 2c_m^{ap}$ and $c_m^{p} =c_m^{pp}$.}
	\label{FigSI_passive_motor_linker_scheme}
\end{figure}

We note the concentration of passive and active motors $c_m^p$ and $c_m^a$, respectively, with $c_m^p +c_m^a= c_m^0$, the concentration of streptavidin $c_{str}^0$ and the concentration of links $c_{l}^0$. In the experiments $c_{str}^0 =c_m^0$, which leads to different types of complexes. To simplify we will consider that we have in same quantities streptavidin with 0,1 or 2 kinesin attached, which seems reasonable considering usual binding equations and values of thermodynamic and kinetic constants \cite{srisa-art_monitoring_2008}. Only 2-kinesin complexes can be considered as linkers. Among these linkers, we have a proportion of $1-\left(\frac{c_m^p}{c_m^0}\right)^2$ active linkers and of $\left(\frac{c_m^p}{c_m^0}\right)^2$ passive linkers (Fig.~S\ref{FigSI_passive_motor_linker_scheme}).
Then the concentration of passive linkers is 
\begin{equation}
	c_l^p = \left( \frac{c_m^p}{c_m^0}\right)^2\frac{c_{str}^0}{3}. \nonumber
\end{equation}

With $c_m^p \approx 20$~nM, as calculated in the MT, and $c_m^0 = 25$~nM we find $c_l^p\approx 5$~nM.

\newpage
\section{Hydrodynamic theory: Instabilities of nematic gels}

In these experiments, a motor-microtubule gel in a thin channel is examined at different ATP concentrations. At high pluronic concentration, the characteristic instability of the gel changes with ATP concentration.
While at high ATP concentrations, the gel has the usual bend instability characteristic of active nematic fluids, at low ATP concentrations, it forms an essentially two-dimensional sheet which undergoes a buckling instability, without any significant in-plane distortion of the nematic alignment. 
It is hypothesised that this spectacularly distinct behaviour at low and high ATP concentrations is a result of an ATP-induced fluidisation transition in the motor-microtubule system. That is, it is hypothesised that at low ATP concentrations, most of the kinesin motors are passive and act as passive crosslinkers resulting in an essentially permanently crosslinked, uniaxial gel. At high ATP concentrations, in contrast, most of the motors are activated and move along the microtubule fibres, thereby fluidising the gel. This hypothesis is experimentally supported by the fact that introducing passive crosslinkers in a high-ATP gel, that in their absence loses orientational order via a bend instability, leads to the formation of a layer which buckles out of plane but retains in-plane orientational order, just as in a low-ATP gel.
In this supplement we demonstrate that the distinct instability modes observed at low and high ATP concentrations are consistent with the hypothesised gelation transition. We adapt a consistent hydrodynamic model of polymeric active uniaxial systems developed in \cite{Ano_poly} to describe a viscoelastic active gel in both the viscous and elastic, permanently-crosslinked regimes. In the viscous regime, the model displays the usual bend instability of the orientational order associated with viscous nematic fluids. In the permanently-crosslinked regime, in contrast, the instability of the ordered state is suppressed \cite{Ano_sol}. Instead, the gel forms a film which buckles while preserving in-plane order, as observed in the experiments. We now recapitulate the model of \cite{Ano_poly}, whichwe will use, in detail for completeness. 

A nematic gel in which the filaments are crosslinked for a finite time $\bar{\tau}_C$ is described in terms of three variables: the nematic order parameter $\bar{{\bsf Q}}$, the conformation tensor $\bar{{\bsf C}}$ and the fluid velocity field ${\bf V}$. The symmetric conformation tensor reduces to the strain tensor of the gel in the limit in which the gel is crosslinked over infinitely long time scales \cite{Ano_poly, Ano_sol}. The coupled hydrodynamic equations for a crosslinked gel, which amounts to an active version of Johnson-Segelman model, were described in \cite{Ano_poly}:
\begin{equation}
	\label{Qeqn3d}
	\partial_t\bar{{\bsf Q}}=\bar{{\bsf Q}}\cdot\bar{\bm{\omega}}-\bar{\bm{\omega}}\cdot\bar{{\bsf Q}}-2\bar{\lambda}\bar{{\bsf A}}+\frac{1}{\bar{\tau}_Q}\bar{{\bsf H}},
\end{equation}
\begin{equation}
	\label{Ceqn3d}
	\partial_t\bar{{\bsf C}}=\bar{{\bsf C}}\cdot\bar{\bm{\omega}}-\bar{\bm{\omega}}\cdot\bar{{\bsf C}}-2\bar{\lambda}_C\bar{{\bsf A}}+\frac{1}{\bar{\tau}_C}\bar{{\bsf B}}
\end{equation}
where $\bar{{\bsf A}}=(1/2)[\nabla{\bf V}+(\nabla{\bf V})^T]$, $\bar{\bm{\omega}}=(1/2)[\nabla{\bf V}-(\nabla{\bf V})^T]$, and ${\bsf H}=-\delta \bar{F}/\delta\bar{{\bsf Q}}$ and ${\bsf B}=-\delta \bar{F}/\delta\bar{{\bsf C}}$ are the molecular fields corresponding to the nematic tensor and the conformation tensor respectively where $\bar{F}$ is the free energy that would control the dynamics in the absence of activity. The equation for the conformation tensor has been simplified relative to \cite{Ano_poly} without, however, changing any of the essential physics. The term with the coefficient $\bar{\lambda}$ in \eqref{Qeqn3d} describes the flow-alignment or how the elongated elements rotate in response to a shear flow. For flow-aligning systems, it has a value $|\bar{\lambda}|>1$ and for flow-tumbling systems, $|\bar{\lambda}|<1$.
The term with the coefficient $\bar{\lambda}_C$ has a somewhat different physics: in the Johnson-Segelman model of polymeric systems, which \eqref{Ceqn3d} reduces to in the absence of any coupling to the orientational order parameter, this describes the slip parameter which accounts for non-affine polymeric deformations. In this case, it accounts for the non-affine deformation of the crosslinked microtubule gel. If the deformation is affine, which is expected to be the case for a permanently crosslinked gel (i.e., when $\bar{\tau}_C\to\infty$), $\bar{\lambda}_C=-1$. When the gel is not permanently crosslinked, and in particular, in the fluid limit $|\bar{\lambda}_C|<1$ and generally, in this limit, $\bar{\lambda}_C<0$. That $\bar{\lambda}_C=-1$ when the gel is permanently crosslinked can be seen from the following argument: in this limit, the fluctuations of the conformation tensor is equivalent to the fluctuations of the strain tensor which should be equal to the strain rate tensor to the lowest order in gradients.

The standard form of the free energy density is
\begin{equation}
	\label{fenrg13d}
	\bar{f} = \left[ \frac{\bar{\alpha}}{2} \mbox{Tr}{\bar{\bsf Q}}^2 + \frac{\bar{\beta}}{4} (\mbox{Tr}\bar{{\bsf Q}}^2)^2 \right] + \frac{\bar{K}}{2}
	(\nabla \bar{{\bsf Q}})^2
	+\frac{1}{2}\mbox{Tr}(\bar{{\bsf C}} -{\bsf I})^2 +
	2\bar{\chi}\bar{{\bsf C}}:\bar{{\bsf Q}} +\bar{\kappa} \mbox{Tr}(\bar{{\bsf C}}-{\bsf I})(\mbox{Tr}\bar{{\bsf Q}}^2)\;,
\end{equation}
with the free energy $\bar{F}=\int_{\bf r} \bar{f}$.

The velocity field, for slow flows relevant to biological experiments is given by the Stokes equation which balances viscous forces with other forces and has the form
\begin{equation}
	\label{vel3d}
	\bar{\eta}\nabla^2{\bf V}=\nabla\bar{\Pi}-\nabla\cdot\bar{\bm{\sigma}}^G
\end{equation}
where $\bar{\Pi}$ is the pressure that enforces three-dimensional incompressibility constraint $\nabla\cdot{\bf V}=0$ and  $\bar{\bm{\sigma}}^G$ is the stress due to the gel which has the form described in \cite{Ano_poly}: 
\begin{equation}
	\label{eqstrs3d}
	\bar{\bm{\sigma}}^G=- \bar{K}(\nabla \bar{{\bsf Q}}):(\nabla\bar{{\bsf Q}})+ 2 ({\bar{{\bsf Q}}\cdot\bar{{\bsf H}}})^A+2\bar{\lambda}\bar{{\bsf H}}+2\bar{\lambda}_c\bar{{\bsf B}}+ 2 ({\bar{{\bsf C}}\cdot\bar{{\bsf B}}})^A-\bar{\zeta}\bar{{\bsf Q}}
\end{equation}
where the term with $\bar{\zeta}$ is the active stress, which we assume to depend only on the nematic order parameter, with $\bar{\zeta}>0$ describing a system with an extensile \emph{uniaxial} active stress. The remaining terms are required to ensure that the correct equilibrium physics is obtained in the absence of activity. 

This model describes the dynamics of a crosslinked active gel. Now we will examine the predictions stemming from this model both in the regime in which the crosslinking time is small i.e., $\bar{\tau}_C\to 0$, and the gel is essentially fluid and when it is large i.e., $\bar{\tau}_C\to \infty$ and the gel is essentially permanently crosslinked, in the geometry appropriate for the experiment.


We first examine the fluid limit.

\subsection{The fluid limit}
The experiment is performed in a channel whose width of the channel is $\sim10$ times the height, the length is $\sim 10$ times the width. We take the shortest dimension of the channel to be along $z$ and denote its thickness by $H$.
For the active gel at high concentrations of ATP and pluronic, the concentration of the gel along $z$ direction appears to be constant and to fill the channel in that direction, at least at the start of the instability. The base state is ordered along $x$ -- the long axis of the channel. To examine the stability of the gel in this geometry, it is useful to average the equations of motion along $z$. Since the channel walls impose no-penetration boundary condition of the velocity field, $V_z$, the $z$ component of the three-dimensional velocity field must vanish upon averaging over the $z$ direction.
The thickness-averaged, effectively two-dimensional velocity field is denoted by ${\bf v}=(1/H)\int_{0}^Hdz{\bf V}_\perp$, where $\perp$ denotes in-plane components of the velocity field. Similarly, we average $\bar{{\bsf Q}}$ and $\bar{{\bsf C}}$ along the $z$ direction and project them in the two-dimensional plane (since the orientational fluctuations in the $z$ direction are expected to average out in the $z$ direction, we expect $1/H\int_{0}^Hdz\bar{Q}_{zz}=1/H\int_{0}^Hdz\bar{Q}_{z\perp}=0$ which is consistent with the gel remaining homogeneous in the $z$ direction) defining ${\bsf Q}=1/H\int_{0}^Hdz\bar{{\bsf Q}}_{\perp\perp}$ and ${\bsf C}=1/H\int_{0}^Hdz\bar{{\bsf C}}_{\perp\perp}$. Similarly, all terms appearing in equations of motion and the free energy are also averaged over the thickness with the averaged quantities being represented by the unbarred version of the corresponding barred, three-dimensional coefficients. See \cite{Ano_apol, Ano_pol} for details.

The equations of motion for ${\bsf Q}$ and ${\bsf C}$ are given by \eqref{Qeqn3d} and \eqref{Ceqn3d} with all the three-dimensional coefficients appearing in those equations being replaced by their $z$-averaged, unbarred counterparts:
\begin{equation}\label{eq_tensorQ}
	\partial_t{\bsf Q}={\bsf Q}\cdot\bm{\omega}-\bm{\omega}\cdot{\bsf Q}-2\lambda{\bsf A}+\frac{1}{\tau_Q}{\bsf H}
\end{equation}
\begin{equation}\label{eq_tensorC}
	\partial_t{\bsf C}={\bsf C}\cdot\bm{\omega}-\bm{\omega}\cdot{\bsf C}-2\lambda_C{\bsf A}+\frac{1}{\tau_C}{\bsf B}
\end{equation}
where ${\bsf A}=(1/2)[\nabla_\perp{\bf v}+(\nabla_\perp{\bf v})^T]$, $\bm{\omega}=(1/2)[\nabla_\perp{\bf v}-(\nabla_\perp{\bf v})^T]$ and the free energy density is 
\begin{equation}
	\label{fenrg1}
	f = \left[ \frac{\alpha}{2} \mbox{Tr}{\bsf Q}^2 + \frac{\beta}{4} (\mbox{Tr}{\bsf Q}^2)^2 \right] + \frac{K}{2}
	(\nabla_\perp {\bsf Q})^2
	+\frac{1}{2}\mbox{Tr}({\bsf C} -{\bsf I})^2 +
	2\chi{\bsf C}:{\bsf Q}\;,
\end{equation}
with $F=\int_{\bf x}f$ where ${\bf x}$ is a two-dimensional vector. Here, for simplicity, we have taken $\kappa$ -- the thickness averaged version of $\bar{\kappa}$ -- to be $0$ for simplicity since it doesn't affect the linear stability of the ordered gel in the fluid regime qualitatively.
From (\ref{fenrg1}), the explicit expressions for the molecular fields are 
\begin{equation}
	{\bsf H} = -\left[\alpha{\bsf Q }+ \beta \mathrm{Tr}{\bsf Q}^2{\bsf Q} \right] +K \nabla_\perp^2 {\bsf Q}
	- 2\chi{\bsf C}^T
\end{equation}
and 
\begin{equation}
	{\bsf B}= -({\bsf C}-{\bsf I}) -2\chi{\bsf Q}.
\end{equation}

The constitutive equation for the fluid velocity is qualitatively modified from \eqref{vel3d} due to the averaging over thickness. Due to the $z$ confinement, which imposes a no-slip boundary condition on the in-plane components of the velocity field, the three-dimensional hydrodynamics is screened at the scale $H$. Averaging $(\bar{\eta}/H)\int_{0}^H dz \partial_z^2{\bf V}_\perp$ over the thickness, we find a frictional screening of the flow with a friction coefficient $\Gamma={\eta}/(12H^2)$. Here, the numerical factor is characteristic of a Poiseuille flow, and other flow profiles will have different numerical factors, but will not modify the scaling of $\Gamma$ with $H$. Therefore, the constitutive equation for the thickness-averaged velocity field is 
\begin{equation}
	\label{eqvel}
	\Gamma{\bf v}-\eta\nabla_\perp^2{\bf v}=-\nabla_\perp\Pi+\nabla_\perp\cdot\bm{\sigma}^G
\end{equation}
where $\Pi$ enforces the \emph{two-dimensional} incompressibility constrain $\nabla_\perp\cdot{\bf v}=0$ and $\bm{\sigma}^G$ is the thickness-averaged version of \eqref{eqstrs3d}:
\begin{equation}
	\label{eqstrs}
	\bm{\sigma}^G=- K(\nabla_\perp {\bsf Q}):(\nabla_\perp{\bsf Q})+ 2 ({{\bsf Q}\cdot{\bsf H}})^A+2\lambda{\bsf H}+2\lambda_c{\bsf B}+2 ({{\bsf C}\cdot{\bsf B}})^A-\zeta{\bsf Q}.
\end{equation}
This is the thickness averaged version of the model presented in \cite{Ano_poly} recapitulated here for completeness.

With this model, we now calculate the linear stability of an effectively two-dimensional gel (after an average over the $z$ direction) aligned along $\hat{x}$ with 
\begin{equation}
	{\bsf Q}^0=S_0\begin{pmatrix}1 &0\\0 &-1\end{pmatrix},
\end{equation}
and ${\bf v}=0$. In this no flow steady state, $\mathrm{Tr}{\bsf C}^0=2$ and ${\bsf C}^0-{\bsf I}=-2\chi{\bsf Q}^0$. This implies that $S_0^2=|\alpha-4\chi^2|/2\beta$ when $\alpha-4\chi^2<0$. We are interested in the stability of a perfectly ordered state with $\alpha\ll0$ and we take $S_0=1$ without loss of generality.

For deviations away from the perfectly ordered state, when $\alpha<0$,
\begin{equation}
	{\bsf H}=|\alpha|\delta{\bsf Q}-\beta\mathrm{Tr}[({\bsf Q}^0)^2+{\bsf Q}_0\cdot\delta{\bsf Q}+\delta{\bsf Q}\cdot{\bsf Q}_0]({\bsf Q}^0+\delta{\bsf Q})+\beta\mathrm{Tr}({\bsf Q}^0)^2{\bsf Q}^0+K\nabla_\perp^2\delta{\bsf Q}-2\chi\delta{\bsf C}^T.
\end{equation}
Expanding 
\begin{equation}
	\delta{\bsf Q}=(1+\delta S)\left[\begin{pmatrix}\cos2\theta &\sin 2\theta\\\sin2\theta &-\cos2\theta\end{pmatrix}-\begin{pmatrix}1 &0\\0 &-1\end{pmatrix}\right],
\end{equation} 
where $\theta$ is the local deviation of the gel orientation away from $\hat{x}$,
we get to linear order in $\delta S$,
\begin{equation}
	H_{xx}=-(2|\alpha|+12\chi^2)\delta S-2\chi\delta C_{xx}^T+K\nabla_\perp^2\delta S
\end{equation}
while to linear order in $\theta$,
\begin{equation}
	H_{xy}=-8\chi^2\theta-2\chi\delta C_{xy}+K\nabla_\perp^2\theta
\end{equation}
while
\begin{equation}
	B_{xx}=-\delta C_{xx}-1-2\chi\delta S
\end{equation}
and
\begin{equation}
	B_{xy}=-\delta C_{xy}-4\chi\theta.
\end{equation}
In the limit of small $\tau_C\to 0$, i.e., when the crosslinking time of the gel is small and therefore it is essentially fluid, the equation for the conformation tensor reduces to ${\bsf B}=0$. This is the limit that we examine in this section. The magnitude of the order parameter
$\delta S$ also has finite zero-wavenumber relaxation rate $2|\alpha+4\chi^2|/\tau_Q$ and relaxes to $0$ in a finite time; i.e., $S$ relaxes to $S_0=1$ in a finite time. In contrast, the relaxation or growth rate of angular fluctuations is $\mathcal{O}(\nabla_\perp^2)$ since $\delta C_{xy}=-4\chi\theta$ and therefore, $-8\chi^2\theta-2\chi\delta C_{xy}=0$. This is expected since in a fluid, the angular fluctuations are the Goldstone modes of the broken rotation symmetry and cannot relax in finite time in an infinite system. In the opposite, elastomeric, limit, which we will examine in greater detail in the next section, when $\tau_C\to\infty$, i.e. the gel is essentially permanently crosslinked \emph{both} the amplitude and angular fluctuations are slaved to the conformation tensor and relax to it in a finite time. This is analogous to the Higgs-Anderson mechanism for a Goldstone mode acquiring a mass and has been discussed in detail in the context of passive and active nematic elastomers \cite{Xing_Lubensky, Ano_sol}. 

Since $\delta S$ fluctuations have a finite, zero-wavenumber relaxation rate, while $\theta$ fluctuations do not, we eliminate $\delta S$ fluctuations and consider a theory purely in terms of the angle field. That is, we consider angular fluctuations about a state with perfect nematic alignment. The linearised equation for the angle field is
\begin{equation}
	\label{eqang}
	\partial_t\theta=\frac{1-\lambda}{2}\partial_xv_y-\frac{1+\lambda}{2}\partial_yv_x-\frac{\chi}{\tau_Q}[\delta C_{xy}+4\chi\theta]+\frac{K}{\tau_Q}\nabla_\perp^2\theta
\end{equation}
which only couples to the $x-y$ component of the conformation tensor directly. The linearised equation of motion for $\delta C_{xy}$ is
\begin{equation}
	\label{eqcxy}
	\partial_t\delta C_{xy}=-\lambda_C(\partial_x v_y+\partial_yv_x)-\frac{1}{\tau_C}[\delta C_{xy}+4\chi\theta].
\end{equation}
The $\mathrm{Tr}[\delta {\bsf C}]$ cannot appear in the dynamical equation via the coupling to the velocity field since, due to incompressibility, the velocity is insensitive to it. Therefore, its relaxation is completely independent and happens with a relaxation rate $1/\tau_C$. However, the deviatoric part of ${\bsf C}$ -- that is  $\delta {\bsf C}^T$, $\delta C_{xx}^T$ does couple to $\delta C_{xy}$ and $\theta$ through the velocity field. This has a linearised equation of motion
\begin{equation}
	\label{eqcxx}
	\partial_t\delta C_{xx}^T=-\lambda_C(\partial_x v_x-\partial_yv_y)-\frac{1}{\tau_C}\delta C_{xx}^T.
\end{equation}
Solving for the velocity field in the Fourier space, for fluctuations with wavevector ${\bf q}_\perp\equiv(q_x, q_y)$, by expanding the stress in \eqref{eqvel}, given by \eqref{eqstrs}, to linear order in fluctuating quantities, eliminating the pressure by using the transverse projector $\mathcal{P}_{ij}=\delta_{ij}-\hat{q}_{\perp_i}\hat{q}_{\perp_j}$, to take incompressibility into account, and injecting this into \eqref{eqang}, \eqref{eqcxy} and \eqref{eqcxx}, we obtain closed linear dynamical equation for $\theta$, $\delta C_{xy}$ and $\delta C_{xx}^T$. We then obtain the eigenvalues of this linear system. At small $\tau_C$, two of the eigenvalues vanish as $-1/\tau_C$ corresponding to the fast relaxation of the two components of $\delta {\bsf C}^T$. The final eigenvalue, which corresponds to the angular fluctuations, is 
\begin{equation}
	\label{eigenfull}
	\Xi_\theta=-q_\perp^2\left[\frac{K}{\tau_Q}-\frac{\zeta\cos2\phi}{\Gamma}(1-\lambda\cos2\phi)\right]+\frac{2\tau_C\chi}{\tau_Q}q_\perp^2\left[\frac{2\chi K}{\tau_Q}-\frac{\zeta\cos2\phi}{\Gamma}\{2\chi-(\lambda\chi+\lambda_C)\cos2\phi\}\right]
\end{equation}
where $\phi$ is the angle between the ordering direction $\hat{x}$ and the wavevector ${\bf q}_\perp$ and $q_\perp$ is the magnitude of the wavevector $q_\perp\equiv|{\bf q}_\perp|$, i.e., $q_x=q_\perp\cos\phi$ and $q_y=q_\perp\sin\phi$. As expected, to $\mathcal{O}(\tau_C^0)$, there is no effect of coupling to the conformation tensor on the angular fluctuations; in this limit the conformation tensor relaxes infinitely fast and the usual mode structure characteristic of an active nematic fluid in a confined geometry is recovered \cite{Voit, Aditi1, RMP}. The modification due to the coupling with the conformation tensor appears only at first order in $\tau_C$. At zeroth order in $\tau_C$, for $\zeta>0$, i.e., for an extensile system, which motor-microtubule fluids are, activity has a destabilising influence on the ordered state for $\phi\lesssim \pi/4$. For $\lambda<1$, this continues all the way to $\phi=0$ i.e., for perfect bend perturbations. In the confined system, this destabilising influence is resisted by the Frank elasticity of the gel, but for large enough $\zeta$, the gel is unstable. In the experiments, the gel has a bend distortion. For $\phi=0$, i.e., for pure bend, the eigenvalue corresponding to the angular fluctuations reduces to 
\begin{equation}
	\label{eq_fluid_bending_inst}
	\Xi_\theta=-q_\perp^2\left[\frac{K}{\tau_Q}-\frac{\zeta}{\Gamma}(1-\lambda)\right]+\frac{2\tau_C\chi}{\tau_Q}q_\perp^2\left[\frac{2\chi K}{\tau_Q}-\frac{\zeta}{\Gamma}\{2\chi-(\lambda\chi+\lambda_C)\}\right]
\end{equation}
This describes a small wavenumber instablity for $\zeta\gg K\Gamma/\tau_Q$, $\lambda<1$ and $\tau_C\to 0$. The large wavenumber modes are stabilised by Frank elasticity entering via the particle phase stress \eqref{eqstrs} \cite{Ano_apol}. This implies that the fastest growing mode just beyond the onset of the instability will have a characteristic wavenumber $\propto \sqrt{\Xi_\theta/q_\perp^2}$ and a lengthscale $\propto 1/ \sqrt{\Xi_\theta/q_\perp^2}$. Assuming that the stabilising fourth order term is essentially independent of activity, the lengthscale should therefore scale as $1/\sqrt{\zeta}$. Further, it should also depend on $\tau_C$. Since upon changing ATP concentrations in the bending regime, no significant change in the pattern wavelength is observed, it implies that ATP concentrations do not significantly modify either the activity or the crosslinking time at least at high concentrations. However, since the gel behaviour changes from low ATP buckling regime to the high ATP bending regime, the crosslinking time must be affected by ATP concentration. We speculate that ATP drives a sharp transition in the crosslinking time $\tau_C$ from a high, essentially infinite, value at low concentration to a small value at a critical value of ATP concentration. Previous studies have also found that motor properties stop changing at high-enough ATP concentrations. 

Note that in the discussion above we have ignored higher order in gradients bulk active stresses that when averaged over the channel thickness lead to an active force of the same order in in-plane gradients as $\nabla_\perp\cdot{\bsf Q}$, but distinct angular symmetry \cite{Ano_apol}. This force can, in principle, stabilise active fluids if it has the right sign. However, since the active fluid in the experiments is not stable, we conclude that this force is either destabilising or small compared to the one retained here.

The theoretical developments in this section have been concerned with the fluid limit in which $\tau_C\to0$. However, \eqref{eqang}, \eqref{eqcxy} and \eqref{eqcxx} are also obviously valid in the opposite limit of $\tau_C\to\infty$ in which the gel is essentially permanently crosslinked. In this limit, one of the three eigenvalues are still $\mathcal{O}(q_\perp^2)$. However, in this limit, the eigenvector corresponding to this eigenvalue is not controlled by angular fluctuations but by $\delta C_{xy}$ fluctuations. The eigenvalue corresponding to angular fluctuations now acquires a finite relaxation rate. This can be understood by examining \eqref{eqang} and \eqref{eqcxy}. In the $\tau_C\to\infty$ limit, there is no relaxation of $\delta C_{xy}$ fluctuations to zeroth order in $\tau_C$ which is instead slaved to the strain rate (as discussed earlier,  $\lambda_C\to-1$ in this limit). Because of this, its relaxation rate is $\sim \mathcal{O}(q_\perp^2)$. From \eqref{eqang}, this further implies that the angular fluctuations must relax fast (i.e. with a wavenumber independent relaxation rate) to a value governed by the conformation tensor. We now display the three eigenvalues in this limit to demonstrate that they can all be stable even for $\zeta>0$.
\begin{equation}
	\Xi_1(\tau_C\to\infty)=-\frac{q_\perp^2}{\Gamma}\lambda_C(\lambda_C+2\lambda\chi)(1-\cos4\phi)-\frac{1}{\tau_C}
\end{equation}
This eigenvalue corresponds to the $\delta C_{xx}$ fluctuations. This is stabilising when $\chi>0$ (as we will see is required for stability and argue is the case in the next section) as $\lambda_C\to -1$ and when $\lambda$ is small and negative or positive. The magnitude of $\lambda$ in motor microtubule experiments is likely to be $|\lambda|<1$ because of the microtubule length. Therefore, this eigenvalue is likely to be stabilising. Also, note that this relaxation rate is independent of activity.
The second eigenvalue, predominantly corresponding to $\delta C_{xy}$ fluctuations is
\begin{equation}
	\label{Eigen2}
	\Xi_2(\tau_C\to\infty)=\frac{q_\perp^2\zeta\lambda_C\cos^22\phi}{2\Gamma\chi}-q_\perp^2\left[\frac{K}{4\tau_C\chi^2}-\frac{\zeta\tau_Q\cos2\phi}{8\Gamma\chi^3\tau_C}\{2\chi-(\lambda_C+2\lambda\chi)\cos2\phi\}\right].
\end{equation}
The first term is stabilising when $\lambda_C\to -1$, $\zeta>0$ and $\chi>0$. The first condition is fulfilled in this essentially permanently corsslinked regime while the second condition simply implies extensility. We will argue in the next section that $\chi>0$ in our system. Therefore, the first $\tau_C$ independent term is stabilising in our system. While it vanishes precisely for $\phi=\pi/4$, the eigenvalue remains stabilising in this case as is evident from the $\mathcal{O}(1/\tau_C)$ term at $\phi=\pi/4$. Finally, the third eigenvalue, primarily corresponding to angular fluctuations is 
\begin{multline}
	\Xi_3(\tau_C\to\infty)=-\frac{4\chi^2}{\tau_Q}-q_\perp^2\left[\frac{K}{\tau_Q}-\frac{\{\zeta+4\chi(\lambda_C+2\lambda\chi)\}\cos2\phi\{2\chi-(\lambda_C+2\lambda\chi)\cos2\phi\}}{2\Gamma\chi}\right]-\frac{1}{\tau_C}\\+q_\perp^2\left[\frac{K}{4\tau_C\chi^2}-\frac{\zeta\tau_Q\cos2\phi\{2\chi-(\lambda_C+2\lambda\chi)\cos2\phi\}}{8\Gamma\tau_C\chi^2}\right].
\end{multline}
This clearly implies that in the limit of almost permanently corsslinked gel, the angular fluctuations have a finite wavenumber-independent decay rate and are stable. 

\begin{figure}
	\centering
	\includegraphics[width=9cm]{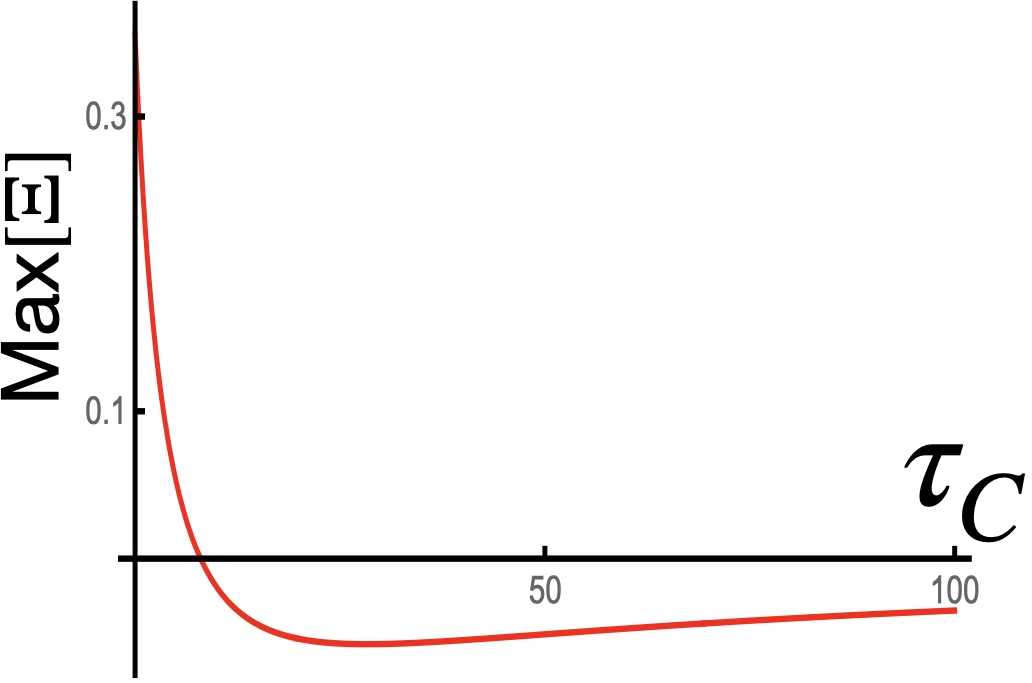}
	\renewcommand{\figurename}{Figure S\!\!}
	\caption{Representative plot of the eigenvalue $\Xi$ at $|q|_\perp=1$, in the direction $\phi$ in which it is maximum, as a function of $\tau_C$. The values of the other parameters are $K=0.2$, $\lambda_C=-1$, $\lambda=0.9$, $\Gamma=1$, $\chi=0.1$, $\tau_Q=1$ and $\zeta=2$. All of these parameters have signs that we expect in our system. Importantly, this demonstrates that as a function of $\tau_C$, the maximum of $\Xi$ goes from a positive value, signifying an unstable mode to a negative one, implying a stable mode. This demonstrates that crosslinking can stabilise a nematic gel.}
	\label{eigenfig}
\end{figure}
More generally, at a finite $\tau_C$, two of the three modes have finite, zero wavenumber relaxation rates. One of these two has the eigenvalue $-1/\tau_C$ and the other $-(1/\tau_C+4\chi^2/\tau_Q)$. Clearly, both of these diverge as $\tau_C\to 0$ as discussed above, while the former vanishes in the $\tau_C\to\infty$ limit and the latter goes to a finite non-zero value. The first eigenvalue has an eigenvector that predominantly corresponds to $\delta C_{xx}$ for all $\tau_C$. However, an examination of the eigenvector corresponding to the second eigenvalue reveals that it predominantly corresponds to $\delta C_{xy}$ fluctuations at small $\tau_C$ but to angular fluctuations at large $\tau_C$. That is, while at small $\tau_C$ conformation tensor fluctuations are non-hydrodynamic and angular fluctuations are hydrodynamic, in the $\tau_C\to\infty$ limit, exactly the opposite happens. However, importantly, both of these eigenvalues are stabilising at small wavenumbers. Therefore, the overall stability of the system depends only on the final eigenvalue which is $\mathcal{O}(q^2)$ and predominantly corresponds to angular fluctuations at small $\tau_C$ and to $\delta C_{xy}$ fluctuations at large $\tau_C$. This eigenvalue is 
\begin{equation}
	\Xi=-\frac{q_\perp^2}{\tau_Q+4\chi^2\tau_C}\left[K-\frac{\tau_Q\zeta}{\Gamma}\cos2\phi\left\{1-\left(\lambda-\frac{2\lambda_C\chi\tau_C}{\tau_Q}\right)\cos2\phi\right\}\right]
\end{equation}
It is clear that this eigenvalue reduces to \eqref{eigenfull} and \eqref{Eigen2} in the limits of $\tau_C\ll \tau_Q$ and $\tau_C\gg\tau_Q$ respectively. More importantly, when $\zeta\gg K$, $\lambda_C\to -1$ and $\chi>0$, as we expect our system parameters to be, this eigenvalue goes from being \emph{destabilising} to \emph{stabilising} as $\tau_C$ is increased (see Fig. \ref{eigenfig}). Also, since the \emph{maximum} value of this eigenvalue, at a fixed $q_\perp$ and for fixed values of other parameters, is positive (i.e. destabilising) for $\tau_C\to0$, is negative for intermediate $\tau_C$ for a finite $K$ and is $0$ in the $\tau_C\to\infty$ limit (since the maximum value of $\Xi_2$ in \eqref{Eigen2} is $0$, for $\phi=\pi/4$ in this limit), it must have a minimum as a function of $\tau_C$, as seen from Fig. \ref{eigenfig}.

This discussion therefore demonstrates that orientational fluctuations of the gel are stabilised by crosslinking. This is consistent with the experimental observation. However, in the experiments, the gel forms a film which does not occupy the full channel in the $z$ direction which then buckles out of the plane. Since in this calculation we have averaged over the entire thickness of the channel, this buckling instability is not accessible within this discussion. In the next section, we consider a gel which is in the fully elastomeric regime, i.e., $\tau_C=\infty$ and forms a film. We eliminate the fast angular fluctuations in favour of strain or displacement fluctuations of the now solid (nematic-elastomeric) gel and the height fluctuations of the film. We show that the in-plane fluctuations of the nematic-elastomeric gel are stabilising, which recapitulates the result here, and that the film spontaneously deforms out of the plane to acquire a buckled shape due to active forcing while retaining in-plane order.

	\subsection{The nematic elastomeric limit}
	At small ATP concentrations, the gel is essentially permanently crosslinked and behaves as a poroelastic solid \cite{gagnon_shear-induced_2020}. In this limit, $\bar{\tau}_C\to\infty$ and tthe conformation tensor no longer relaxes.
	Instead, $\bar{{\bsf C}}-{\bsf I}$ becomes the left Cauchy-Green strain tensor of the permanently crosslinked gel \cite{Ano_poly}. The physics of this nematic elastomeric regime of the gel was comprehensively examined in \cite{Ano_sol}. Here we recapitulate the active nematic elastomeric model, for completeness, and show that the predictions of the model are consistent with the experimental observations.
	As in \cite{Ano_sol}, we define $\bar{{\bsf C}}-{\bsf I}=\bar{{\bsf U}}$ where $\bar{{\bsf U}}$ is the strain tensor of the three-dimensional gel. With this identification, the free energy density \eqref{fenrg13d} of the gel transforms to 
	\begin{equation}
		\label{fenrg2}
		\bar{f} = \left[ \frac{\bar{\alpha}}{2} \mbox{Tr}\bar{{\bsf Q}}^2 + \frac{\bar{\beta}}{4} (\mbox{Tr}\bar{{\bsf Q}}^2)^2 \right] + \frac{\bar{K}}{2}
		(\nabla \bar{{\bsf Q}})^2
		+\frac{1}{2}\mbox{Tr}(\bar{{\bsf U}}^T)^2+\frac{\bar{\nu}}{2}(\mbox{Tr}\bar{{\bsf U}})^2 +\bar{\kappa}\mbox{Tr}\bar{{\bsf U}}(\mbox{Tr}\bar{{\bsf Q}}^2)+
		2\bar{\chi}\bar{{\bsf U}}^T:\bar{{\bsf Q}}\;,
	\end{equation}
	were we have introduced a distinct elastic constant $\bar{\nu}$ for $(\mbox{Tr}\bar{{\bsf U}})^2$, which corresponds to the bulk modulus of the gel. In \eqref{fenrg13d}, $\bar{{\bsf C}}-{\bsf I}$ was characterised by a single elastic constant, taken to be $1$, which here would amount to a gel with an equal value for bulk and shear modulus, and that limit can be recovered by setting $\bar{\nu}\to 1$. Since in a permanently crosslinked gel, the nematic order parameter fully relaxes to the strain tensor and modifies its isotropic and deviatoric parts independently, we believe it is clearer to start with a gel with, in principle, distinct coefficients for bulk and shear modulus. 
	
	Rewriting the free energy density in a more convenient form, we get
	\begin{equation}
		\label{fenrg2}
		f_{E} = \left[ \frac{\bar{\alpha}'}{2} \mbox{Tr}\bar{{\bsf Q}}^2 + \frac{\bar{\beta}'}{4} (\mbox{Tr}\bar{{\bsf Q}}^2)^2 \right] + \frac{\bar{K}}{2}
		(\nabla \bar{{\bsf Q}})^2
		+\frac{1}{2}\mbox{Tr}(\bar{{\bsf U}}^T+2\bar{\chi}\bar{{\bsf Q}})^2+\frac{\bar{\nu}}{2}\left(\mbox{Tr}\bar{{\bsf U}}+\frac{\bar{\kappa}}{\bar{\nu}}\mbox{Tr}\bar{{\bsf Q}}^2\right)^2 
	\end{equation}
	where $\bar{\alpha}'=\bar{\alpha}-4\bar{\chi}^2$ and $\bar{\beta}'=\bar{\beta}-2\bar{\kappa}^2/\bar{\nu}$. This implies a homogeneously ordered state with zero velocity with 
	\begin{equation}
		\label{isostrn}
		\mbox{Tr}\bar{{\bsf U}}^0=-\frac{\bar{\kappa}}{\bar{\nu}}\mbox{Tr}\left[{\bar{{\bsf Q}}^{0^2}}\right]
	\end{equation}
	\begin{equation}
		\label{devistrn}
		(\bar{{\bsf U}}^T)^0=-2\bar{\chi}\bar{{\bsf Q}}^0
	\end{equation}
	with the magnitude of the steady-state nematic order parameter being $\bar{S}_0=\sqrt{\bar{\alpha}'/\bar{\beta}'}$, which as earlier, we will take to be $1$ in deeply ordered phase. Note that these strains do not relax at a finite time. This implies that because of nematic alignment, a gel acquires both an isotropic and a uniaxial prestrain. Because $\mbox{Tr}\bar{{\bsf U}}^0=-(\bar{\kappa}/\bar{\nu})(\bar{S}_0)^2$, the gel compresses \emph{isotropically} when $\bar{\kappa}>0$. While this compression is difficult to measure in the low ATP concentration, high pluronic concentration state, since the gel buckles out of plane, it is clearly observed in the low ATP concentration, low pluronic concentration state. We believe that the signs of coefficients do not change upon changing pluronic concentration, which affects the bundling and ordering of microtubules, though their magnitudes may change. Therefore, we believe that $\bar{\kappa}>0$ even in the state with high pluronic concentration. This is further supported by the fact that the gel essentially moves away from the edges of the channel 
	and forms a floating layer in the middle of the channel (though the thickness of the layer is comparable to the channel height, the fact that the gel buckles as is clearly seen from the $xz$ cross-section figures, shows that the gel moves away from the edges of the channel). 
	Therefore, in the steady state, the gel forms essentially a nematic elastomeric sheet because of isotropic contraction.
	
	In addition to the isotropic contraction, there is also a uniaxial contraction \eqref{devistrn}. Again, this is difficult to measure in the state with high pluronic concentration, but in the state with low pluronic concentration, the gel contracts more along the $x$ direction than along the $y$ direction. Therefore, we expect $\bar{\chi}>0$ in our system which means that the gel contracts more along the uniaxial axis, again assuming that the sign of $\bar{\chi}$ doesn't change with pluronic concentration.
	
	Since the gel forms a (thick) film in the channel, we average over the thickness of the gel (\emph{not} the channel), as earlier defining ${\bsf Q}$ and ${\bsf U}$ as the effective two-dimensional orientational order parameter and strain field respectively. As in the last section, the unbarred parameters will correspond to the thickness-averaged (over the gel thickness) versions of the three-dimensional, barred parameters. We take this flat  nematic elastomeric sheet to be situated in the $z=0$ plane.
	Crucially, the three-dimensional velocity field in the channel is not averaged -- with the ${\bsf Q}$ and ${\bsf U}$ tensors being affected by ${\bf V}|_{z=0}$ -- and the ${\bsf Q}$ tensor is \emph{not} projected into the two dimensional plane implying that the nematic director can fluctuate both in the $x-y$ plane as well as in the $z$ direction in this description. 
	
	We now consider fluctuations of this nematic elastomeric sheet. The out-of-plane fluctuations of the sheet is parametrised using the Monge gauge in which a  point on the interface is parametrised by the three-dimensional position vector ${\bf R}=(x, y, h(x,y))$. That is, the displacement of a point on the sheet in the $z$ direction is parametrised by $h(x,y)$.
	%
	With this parametrisation, the normal to the interface is 
	\begin{equation}
		{\bf N}=\frac{\hat{z}-\nabla_\perp h}{\sqrt{1+(\nabla_\perp h)^2}}. 
	\end{equation}
	Since the apolar order parameter can now fluctuate in three dimensions, ${\bsf Q}\equiv S({\bf nn})^T$ where ${\bf n}$ is the director, which is a unit vector. In the steady-state ${\bf n}_0=(1, 0,0)$. The fluctuating director can be written as
	\begin{equation}
		{\bf n}=\frac{\hat{x}+\delta{\bf n}}{\sqrt{1+\delta{\bf n}\cdot\delta{\bf n}}}.
	\end{equation}
	The nematic director is confined to the tangent plane of the nematic elastomer film. Therefore,
	\begin{equation}
		{\bf n}\cdot{\bf N}=0=\frac{1}{\sqrt{1+\delta{\bf n}\cdot\delta{\bf n}}}\frac{1}{\sqrt{1+(\nabla_\perp h)^2}}[\delta n_z-\delta n_y\partial_y h-(1+\delta n_x)\partial_x h] \implies \delta n_z \approx \partial_x h
	\end{equation}
	where the final approximate equality is obtained by retaining only the linear terms. This implies that the fluctuations of the director along $z$ is slaved to the height field. Therefore, the Frank elasticity of the nematic order parameter leads to a bending elasticity $\propto (\partial_x^2h)^2$. We also include an extra bending energy density for the film $f_B=(K_B/2)(\nabla_\perp^2h)^2$, where $\nabla_\perp\equiv(\partial_x,\partial_y)$ denotes an in-plane gradient. We also include a surface tension term $f_S=(\varsigma/2)(\nabla_\perp h)^2$ (a surface tension term is not allowed for a film in a space with three-dimensional rotation invariance, but is allowed in this geometry). Finally, since the film is confined with a channel of thickness $\tilde{H}$, where $\tilde{H}$ is the thickness of the channel minus that of the film itself, the height fluctuations cannot be larger than that. To account for this constraint, we introduce a confining potential for the height field $f_{C}=(\gamma/2) h^2$ whose value is chosen to ensure that height fluctuations are confined within the channel. Therefore, the full free energy for the nematic elastomeric sheet is $F=\int[f_{E}+f_B+f_S+f_C]$. 
	We will now demonstrate that, with this free energy, both the magnitude of the nematic order parameter as well as the director fluctuations along $y$, $\delta n_y\equiv\theta$ are slaved to the strain field. 
	
	The strain field of the nematic elastomeric film, which enters $f_E$ \eqref{fenrg2} can be expressed in terms of a displacement fields $\tilde{\bf u}$ with respect to an isotropic, unprestrained reference state (not the elastomeric steady state; the steady state, with this definition of the strain tensor has prestrains \eqref{isostrn} and \eqref{devistrn}). The full form of the strain depends on both $\tilde{u}$ and the height field $h$:
	\begin{equation}
		U_{ij}=\frac{1}{2}(\partial_i \tilde{u}_j+\partial_j \tilde{u}_i+\partial_i \tilde{u}_k\partial_j \tilde{u}_k+\partial_i h\partial_j h).
	\end{equation}
	However, the dependence of the strain tensor on the $h$ field is only at the nonlinear level and will not affect the linear calculation performed here. In fact, at the linear level, the height field turns out to not couple with any of the in-plane fields such as $\theta$ or $\tilde{\bf u}$. 
	
	The fluctuations in the magnitude of the nematic order parameter relaxes in a finite time to a value determined by the strain tensor
	\begin{equation}
		\delta S=-\frac{\nu}{w}\left(\chi\delta U_{xx}^T+\kappa\mbox{Tr}\delta{\bsf U}\right),
	\end{equation}
	where $w=|\alpha'|+2(\kappa^2/\nu+\chi^2)$ and the in-plane angle field, that is the $y$ component of the director field, relaxes to 
	\begin{equation}
		\theta=-\frac{1}{4\chi}\delta U_{xy},
	\end{equation}
	consistent with the discussion in the last section. 
	
	At this stage, it is convenient to eliminate $\delta S$ and $\theta$ fluctuations in $f_E$ and to transform to \emph{new} displacement variables ${\bf u}$ and new strain tensor $\bm{\eta}$ which are defined with respect to the \emph{elastomeric} steady state and not an isotropic one. This standard procedure, usual in the nematic elastomer literature \cite{Xing_Lubensky, Ano_sol}, leads to the new form of $f_E$ in terms of $\bm{\eta}$:
	\begin{equation}
		\label{fenrgeta}
		f_E=\frac{1}{2}\left[ B_1\eta_{xx}^2+B_2\eta_{yy}^2+B_3\eta_{xx}\eta_{yy}+B_4[\eta_{xy}-\mu(\theta-\Omega)]^2\right]
	\end{equation}
	where $\Omega=(1/2)(\partial_x u_y-\partial_y u_x)$ is the rotation angle of the gel and
	\begin{equation}
		B_1=\frac{\Lambda_\parallel^4}{{w}}\left[\frac{\kappa^2}{\nu}+|\alpha'|\left(\frac{1}{2}+\nu\right)+4\kappa\chi+2\nu\chi^2\right]
	\end{equation}
	\begin{equation}
		B_2=\frac{\Lambda_\perp^4}{{w}}\left[\frac{\kappa^2}{\nu}+|\alpha'|\left(\frac{1}{2}+\nu\right)-4\kappa\chi+2\nu\chi^2\right]
	\end{equation}
	\begin{equation}
		B_3=\frac{2\Lambda_\perp^2\Lambda_\parallel^2}{{w}}\left[\frac{\kappa^2}{\nu}+|\alpha'|\left(-\frac{1}{2}+\nu\right)+2\nu\chi^2\right]
	\end{equation}
	\begin{equation}
		B_4=\frac{(\mathcal{R}+1)\Lambda_\perp^4}{2}.
	\end{equation}
	with $\Lambda_\parallel^2=1+\kappa/2\nu-4\chi$ and $\Lambda_\perp^2=1+\kappa/2\nu+4\chi$ which yields $\Lambda_\parallel^2-\Lambda_\perp^2=-8\chi$ which is a measure of anisotropy of the film. Since we argued that $\chi$ is positive in our system, $\Lambda_\parallel^2-\Lambda_\perp^2<0$ which implies that the contraction along the ordering direction is greater than the direction transverse to it. $\mathcal{R}=\Lambda_\parallel^2/ \Lambda_\perp^2$ and $\mu=(\mathcal{R}-1)/(\mathcal{R}-1)$. Note that upon integrating out $\theta$, the elastic coefficient for $\eta_{xy}$ shears also vanish. This is required by rotation symmetry: a global, in-plane rotation of the nematic ordering direction can be compensated by a deformation \cite{Xing_Lubensky}.
	
	In terms of the new strain fields, the order parameter magnitude fluctuations are slaved to $\bm{\eta}$ as  
	\begin{equation}
		\delta S=\frac{1}{{w}}[(\kappa-\chi/2)\Lambda_\parallel^2\eta_{xx}-(\kappa+\chi/2)\Lambda_\perp^2\eta_{yy}]
	\end{equation}
	and the angle field as 
	\begin{equation}
		\theta=\Omega+\mu^{-1}\eta_{xy}.
	\end{equation}
	Thus all components of the order parameter ${\bsf Q}$ are either slaved to ${\bf u}=(u_x, u_y)$ or to $h$. This implies that all components of the active force ${\bf f}_a=\nabla_\perp\cdot{\bsf Q}\delta(z)$ in the nematic elastomeric film can be expressed in terms of ${\bf u}$ \cite{Ano_sol} and $h$. Doing this and noting that the linearised passive forces are
	\begin{equation}
		{\bf f}_{p}=-\frac{\delta F}{\delta{\bf u}}\delta (z)-\frac{\delta F}{\delta h}\hat{z}\delta(z),
	\end{equation}
	we obtain the force densities due to the gel ${\bf f}^G=\nabla_\perp\cdot\bm{\sigma}^G$ (see \cite{Ano_sol} for details of obtaining ${\bf f}^G$ from the thickness averaged version of $\bar{\bm{\sigma}}^G$ in \eqref{eqstrs3d}):
	\begin{equation}
		\label{fxeq}
		f^G_x=\delta(z)[b_1\partial_x^2u_x+b_2\partial_y^2u_x+b_3\partial_x\partial_yu_y],
	\end{equation}
	\begin{equation}
		\label{fyeq}
		f^G_y=\delta(z)[b_4\partial_x^2u_y+b_5\partial_y^2u_y+b_6\partial_x\partial_yu_x],
	\end{equation}
	\begin{equation}
		\label{fzeq}
		f^G_z=\delta(z)[\varsigma\nabla_\perp^2h-\zeta\partial_x^2h-K_B\nabla_\perp^4h-{K}\partial_x^4h-\gamma h],
	\end{equation}
	where \cite{Ano_sol}
		\begin{equation}
			\label{nu1Q}
			b_1=\left[B_1+\frac{\zeta(-2\kappa+\chi)}{2w}\Lambda_\parallel^2\right],
		\end{equation}
		\begin{equation}
			\label{nu2Q}
			b_2=-\zeta(\beta^{-1}-1),
		\end{equation}
		\begin{equation}
			\label{nu3Q}
			b_3=\frac{1}{2}\left[B_3-2\zeta(\beta^{-1}+1)+\frac{\zeta(2\kappa+\chi)\Lambda_\perp^2}{2{w}}\right],
		\end{equation}
		\begin{equation}
			\label{nu4Q}
			b_4=-\zeta(\beta^{-1}+1),
		\end{equation}
		\begin{equation}
			\label{nu5Q}
			b_5=\left[B_2-\frac{\zeta(2\kappa+\chi)}{2w}\Lambda_\perp^2\right],
		\end{equation}
		\begin{equation}
			\label{nu6Q}
			b_6=\frac{1}{2}\left[B_3-2\zeta(\beta^{-1}-1)-\frac{\zeta(-2\kappa+\chi)\Lambda_\parallel^2}{2{w}}\right].
		\end{equation}
		Note that $b_2$ and $b_4$ are both purely active. This is due to a combination of rotation invariance and time-reversal symmetry \cite{Xing_Lubensky, Ano_sol, Olmsted}. In a passive nematic elastomer, these coefficients would be $0$ since there is no harmonic term in $\eta_{xy}$ in \eqref{fenrgeta} after integrating out $\theta$. This implies that a passive nematic elastomer would be soft for in-plane $x-y$ shears, but an active elastomer acquires a resistance (or is unstable) to such shears \cite{Ano_sol}.
		
		The equation for the fluid velocity is 
		\begin{equation}
			\label{veleq3d}
			\bar{\eta}(\partial_z^2+\nabla_\perp^2){\bf V}=\nabla\bar{\Pi}-{\bf f}^G
		\end{equation}
		where $\bar{\Pi}$ again enforces the three-dimensional incompressibility constraint $\nabla\cdot{\bf V}=0$ and
		which has to be solved in a channel of height $\tilde{H}$. The linearised equation of motion for the in-plane displacement and height fields are simply $\dot{u}_x=V_x|_{z=0}$, $\dot{u}_y=V_y|_{z=0}$ and $\dot{h}=V_z|_{z=0}$, where we have assumed an impermeable and permanently crosslinked film. The constraint of impermeability can be easily relaxed \cite{senoussi_tunable_2019}, but doesn't modify the results qualitatively. To calculate the velocity field at $z=0$, we Fourier transform \eqref{veleq3d} with the wavevector ${\bf q}\equiv(q_x,q_y,q_z)\equiv({\bf q}_\perp, q_z)$, eliminate the pressure $\bar{\Pi}$, which imposes the \emph{three-dimensional} incompressibility constraint $\nabla\cdot{\bf V}=0$, using the three-dimensional transverse projector $\mathcal{P}_{ij}\delta_{ij}-\hat{q}_i\hat{q}_j$ and integrate the velocity fluctuations over $q_z\in\{-\infty,-2\pi/\tilde{H}\}\cup\{2\pi/\tilde{H},\infty\}$. This yields
		\begin{equation}
			\dot{u}_x=\frac{2\tilde{f}_x(1+\sin^2\phi)-f_y\sin2\phi}{8\bar{\eta}|q_\perp|}-\tan^{-1}\left(\frac{\pi}{\tilde{H}|q_\perp|}\right)\frac{2\tilde{f}_x(1+\sin^2\phi)-f_y\sin2\phi}{4\pi\bar{\eta}|q_\perp|}+\frac{\tilde{H}\cos\phi(\tilde{f}_x\cos\phi+\tilde{f}_y\sin\phi)}{\bar{\eta}(4\pi^2+\tilde{H}^2q_\perp^2)}
		\end{equation}
		\begin{equation}
			\dot{u}_y=\frac{2\tilde{f}_y(1+\sin^2\phi)-f_x\sin2\phi}{8\bar{\eta}|q_\perp|}-\tan^{-1}\left(\frac{\pi}{\tilde{H}|q_\perp|}\right)\frac{2\tilde{f}_y(1+\sin^2\phi)-f_x\sin2\phi}{4\pi\bar{\eta}|q_\perp|}+\frac{\tilde{H}\sin\phi(\tilde{f}_x\cos\phi+\tilde{f}_y\sin\phi)}{\bar{\eta}(4\pi^2+\tilde{H}^2q_\perp^2)}
		\end{equation}
		\begin{equation}
			\dot{h}=\tilde{f}_z\left[\frac{1}{4\bar{\eta}|q_\perp|}-\frac{2}{2\bar{\eta}\pi}\tan^{-1}\left(\frac{2\pi}{\tilde{H}|q_\perp|}\right)-\frac{\tilde{H}}{\bar{\eta}(4\pi^2+\tilde{H}^2q_\perp^2)}\right]
		\end{equation}
		where $\phi$ is the angle that ${\bf q}_\perp$ makes with $\hat{x}$ and $\tilde{\bf f}$ are the Fourier transform of the forces in \eqref{fxeq}, \eqref{fyeq} and \eqref{fzeq}:
		\begin{equation}
			\tilde{f}_x=-q_\perp^2(b_1\cos^2\phi u_x+b_2\sin^2\phi u_x+b_3\cos\phi\sin\phi u_y)
		\end{equation}
		\begin{equation}
			\tilde{f}_y=-q_\perp^2(b_4\cos^2\phi u_y+b_5\sin^2\phi u_y+b_6\cos\phi\sin\phi u_x)
		\end{equation}
		and 
		\begin{equation}
			\tilde{f}_z=-[\varsigma q_\perp^2-\zeta q_\perp^2\cos^2\phi-K_B q_\perp^4 -{K}q_\perp^4\cos^4\phi-\gamma]h.
		\end{equation}
		As discussed earlier, there is no coupling between the in-plane displacement modes and the height field at this, linear level. Therefore, they can be analysed independently. At long wavelengths, the eigenvalues for the in-plane modes are
		\begin{equation}
			\label{eq_solid_no_bending_inst}
			\Xi_{\pm}=-\frac{\tilde{H}q_\perp^2}{8\pi^2\bar{\eta}}\left[(b_1+b_4)\cos^2\phi+(b_2+b_5)\sin^2\phi\pm\sqrt{(b_1-b_4)\cos^2\phi+(b_2-b_5)\sin^2\phi-4b_3b_6\sin^22\phi}\right].
		\end{equation}
		As discussed in \cite{Ano_sol}, these modes are not generically unstable for $\zeta>0$ when $\chi>0$. The best way of examining is to concentrate on $\phi=0$ (pure bend) and $\phi=\pi/2$ (pure splay) modes which in the passive elastomer are soft. For these modes, the $u_x$ and $u_y$ equations decouple and the eigenvalues are simply $-[\tilde{H}q_\perp^2/(8\pi^2\bar{\eta})](-b_1,-b_4)$ when $\phi=0$ and $-[\tilde{H}q_\perp^2/(8\pi^2\bar{\eta})](-b_2,-b_5)$ when $\phi=\pi/2$. Both of these modes can be stable when $b_2$ and $b_5$ are greater than $0$ and, in particular, $b_2>0$ implies that a bend instability does not occur. Both $b_2,b_5>0$ when $\zeta>0$ and $\chi>0$. When $\chi>0$, $-1<\beta<0$ and therefore, $\beta^{-1}<-1$. This implies both $\beta^{-1}-1$ and $\beta^{-1}+1$ are negative. Since $\zeta>0$, $b_2=-\zeta(\beta^{-1}-1)$ and $b_5=-\zeta(\beta^{-1}+1)$ are both positive and stabilising. Therefore, the elasticity of the nematic elastomer stabilises against an in-plane instability. Of course, at high enough $\zeta$, $b_1$ and $b_5$ may go unstable, which is analogous to the instability of an active smectic to large extensile active stress \cite{Tapan_smec}, but that requires overcoming the elasticity of the gel which is expected to be large.
		
		
		The decoupled height fluctuations are also affected by activity. In fact, they are essentially the same as in \cite{senoussi_tunable_2019} (which concerned a different experimental regime): when $\zeta>\varsigma$, a band of wavevectors between
		\begin{equation}
			\label{unstab}
			q_{x_\pm}^2=\frac{{\zeta}-\varsigma\pm\sqrt{({\zeta}-\varsigma)^2-4{K}\gamma}}{2}
		\end{equation}
		along $\phi=0$ will be unstable leading to the undulated pattern, where $\bar{K}=K_B+\tilde{K}$. Note that the instability here is not long wavelength, i.e., the film is stable for $q_x\to 0$, due to confinement. The fastest growing mode is again equivalent to the one described in \cite{senoussi_tunable_2019} and has the same scaling with activity. In both $\tilde{H}q_x\to\infty$ and $\tilde{H}q_x\to 0$ limit (the latter is the relevant limit for the experiments since, because the gel occupies almost the entire channel, $\tilde{H}$ is small), this is
		\begin{equation}
			\label{eq_solid_buckling_inst}
			q^*_x=\sqrt{\frac{{\zeta}-\varsigma}{2{K}}}.
		\end{equation}
		When the flat state of the film is unstable towards a patterned conformation, but the in-plane modes are stable, as is the case for $\chi>0$ and $\zeta>0$, the film remains uniaxial in the plane. That is, the microtubules remain oriented in the plane but the film, as a whole, buckles out of the plane, just as is observed in the experiment.
		
		To summarise, we have therefore demonstrated that when the active nematic gel goes from an uncrosslinked, effectively fluid state to a crosslinked essentially solid state, the mode of its instability changes dramatically: from an in-plane bend instability to an out of plane buckling of the contracted layer. Therefore, this discussion supports the hypothesis that the change of the mode of instability of motor-microtubule gels upon changing ATP concentration is due to a gelation or fluidisation transition.
		
		\subsection{Deformation of a nematic fluid film suspended in the middle of the channel}
		In the experiments with the caged ATP, a nematic film is allowed to be formed by passive depletion forces with caged ATP (i.e., at essentially $0$ ATP concentration ). Then the ATP is uncaged and it is at high concentration. This implies that $\tau_C\to 0$ in this state. In this section, we demonstrate how a nematic film floating at $z=0$ in the channel will evolve due to activity in the $\tau_C=0$ limit. In this limit the conformation tensor ${\bsf C}$ relaxes infinitely fast, so we ignore it and consider only nematic fluctuations. We do this in terms of the director ${\bf n}$ with ${\bsf Q}={\bf nn}-(1/3){\bsf I}$.
		For a film initially at $z=0$ with the nematic director aligned along $\hat{x}$, the linearised evolution equations of the fluctuations of the director along $y$ and $z$ directions, $\delta n_y$ and $\delta n_z$ are
		\begin{equation}
			\partial_t \delta n_y=\frac{1-\lambda}{2}\left(\partial_x V_y\right)|_{z=0}-\frac{1+\lambda}{2}\left(\partial_y V_x\right)|_{z=0}+\frac{K}{\tau_Q}\nabla_\perp^2\delta n_y
		\end{equation}
		and
		\begin{equation}
			\partial_t \delta n_z=\frac{1-\lambda}{2}\left(\partial_x V_z\right)|_{z=0}-\frac{1+\lambda}{2}\left(\partial_z V_x\right)|_{z=0}+\frac{K}{\tau_Q}\nabla_\perp^2\delta n_z
		\end{equation}
		where $\nabla_\perp\equiv(\partial_x,\partial_y)$. The force balance equation is 
		\begin{equation}
			\bar{\eta}\nabla^2{\bf V}=\nabla\bar{\Pi}+\delta(z)\zeta\left(\partial_x \delta n_y\hat{y}+\partial_x\delta n_z\hat{z}+\partial_y\delta n_y\hat{x}\right)
		\end{equation}
		where ${\bf V}$ is the three-dimensional velocity field and $\bar{\Pi}$ is the three-dimensional pressure enforcing the incompressibility constraint $\nabla\cdot{\bf V}=0$. These equations are obtained by expanding \eqref{Qeqn3d} and \eqref{vel3d}, with \eqref{eqstrs3d} in the limit in which ${\bsf C}$ relaxes infinitely fast and with the initial state described earlier. Solving for the velocity field in a channel of height $\tilde{H}$, where $\tilde{H}$ is the height of the channel minus the gel thickness, we find that the eigenvalues for \emph{both} $\delta n_y$ and $\delta n_z$ fluctuations are
		%
		%
		%
		%
		%
		%
		%
		%
		%
		%
		%
		\begin{multline}
			\label{eigenflfull}
			\Xi_n=-\frac{Kq_\perp^2}{\tau_Q}-\frac{\zeta |q_\perp|}{4\pi\bar{\eta}}\tan^{-1}\left(\frac{2\pi}{\tilde{H}|q_\perp|}\right)[2\cos2\phi-\lambda(1+\cos^22\phi)]
			\\
			+\frac{\zeta|q_\perp|}{4\bar{\eta}}\left[\cos2\phi-\frac{\lambda}{2}\left(1+\cos^22\phi+\frac{4\tilde{H}|q_\perp|\sin^22\phi}{4\pi^2+\tilde{H}^2q_\perp^2}\right)\right],
		\end{multline}
		where $\phi$ is the angle between the wavevector of perturbation and $\hat{x}$.
		In the limit of $|q_\perp|\gg 1/\tilde{H}$, i.e., for perturbations with in-plane scales \emph{smaller} than the thickness of the channel, this reduces to 
		\begin{equation}
			\label{neflfilm}
			\Xi_n(H\to\infty)=-\frac{Kq_\perp^2}{\tau_Q}+\frac{\zeta|q_\perp|}{8\bar{\eta}}[2\cos2\phi-\lambda(1+\cos^22\phi)]+\frac{\zeta\cos2\phi(1-\lambda\cos2\phi)}{\bar{\eta} \tilde{H}}+\mathcal{O}\left(\frac{1}{\tilde{H}^3}\right).
		\end{equation}
		Note that the angular form of the leading order term in $\tilde{H}$ is distinct from the usual instability of active nematics \cite{Aditi1}. This is due to the \emph{lack of} incompressibility in the plane of the layer i.e., $\nabla_\perp\cdot{\bf V}_\perp|_{z=0}\neq 0$ because $\partial_zV_z|_{z=0}\neq 0$. This should be compared with \eqref{eigenfull} where the nematic fluid filled the channel and, upon averaging over the thickness of the channel, the effective two-dimensional velocity field was incompressible. The usual angular form of the eigenvalue associated with generic active nematic instability only appear at the subleading order in $\tilde{H}$ and is immaterial in the limit $|q_\perp|\gg1/\tilde{H}$. In fact, the leading order in $\tilde{H}$ term is \emph{not}  destabilising when $|\lambda|>1$ and $\lambda\zeta>0$ \cite{Ano_stab}. However, microtubule filaments are likely to be flow-tumbling i.e., $|\lambda|<1$. Therefore, at large $\zeta$, \eqref{neflfilm} signifies an activity-induced \emph{growth} of \emph{both} $\delta n_y$ and $\delta n_z$ fluctuations. For extensile systems, i.e., $\zeta>0$, this growth happens for bend perturbations, i.e., $\phi\approx 0$ for \emph{any} value of $\zeta$. 
		For large in-plane scales, which is the relevant limit for the experiments, i.e., for $|q_\perp|\ll 1/\tilde{H}$, \eqref{eigenflfull} becomes
		\begin{equation}
			\label{eq_fluid_film_bending_and_buckling_inst}
			\Xi_n(|q_\perp|\to 0)=-\frac{Kq_\perp^2}{\tau_Q}+\frac{\zeta \tilde{H}q_\perp^2}{4\pi^2\bar{\eta}}(\cos2\phi-\lambda)
		\end{equation}
		For $|\lambda|<1$, this implies a bend instability of both $\delta n_y$ and $\delta n_z$ fluctuations when $\zeta>4\pi^2 K\eta/(\tau_Q \tilde{H})$. Again, the bend instability would have been avoided for $|\lambda|>1$, because the nematic film is not incompressible in the plane of the film \cite{Ano_stab}.
		
		Since the height fluctuations are slaved to $\delta n_z$ fluctuations as $\delta n_z\approx \partial_x h$, \eqref{eigenflfull} or its limiting versions, \eqref{neflfilm} and \eqref{eq_fluid_film_bending_and_buckling_inst} imply a corresponding instability of the flat conformation of the film with the same eigenvalue. This implies that when a nematic fluid film is activated with high ATP concentrations (such that $\tau_C\to 0$), the director bends in the plane and the flat conformation of the film is destabilised as is observed in the experiments.

\newpage
\section{Supplementary movies}

\setcounter{figure}{0}

\subsection{Bending}

\begin{figure}[H]
	\includegraphics[width=0.8\textwidth]{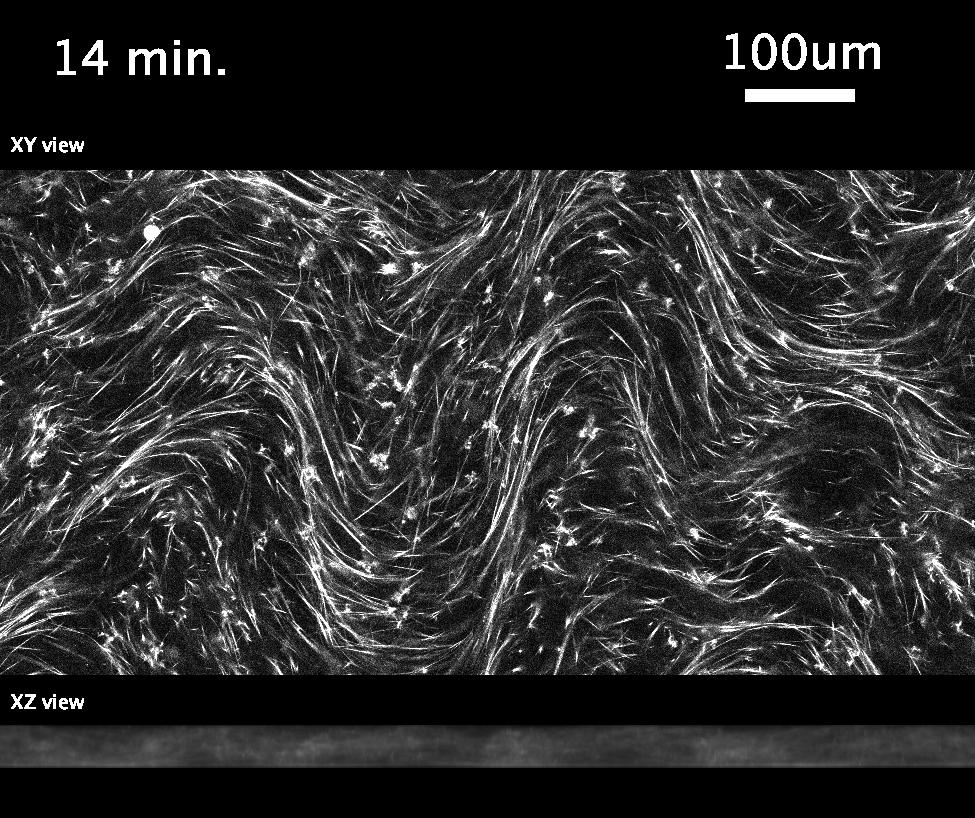}
	\renewcommand{\figurename}{Movie S\!\!}
	\caption{Time-lapse confocal microscopy of microtubules displaying a bending instability. Associated to Figure 1 in the main text. 5.5\% pluronic, [ATP] = 50~$\mu$M.}
	\label{Mov_Fig1_Bending}
\end{figure}

\begin{figure}[H]
	\includegraphics[width=0.5\textwidth]{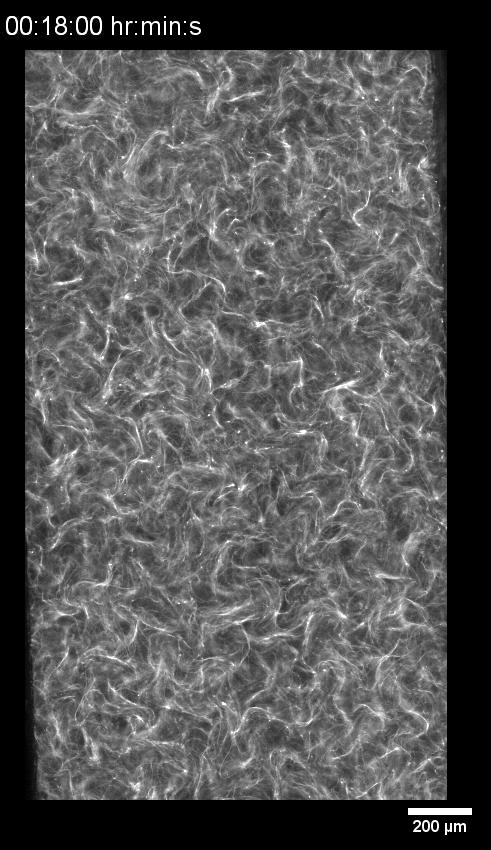}
	\renewcommand{\figurename}{Movie S\!\!}
	\caption{Time-lapse epifluorescence microscopy showing the long-time behavior of microtubules displaying a bending instability. Within 10 minutes, this instability turns into chaotic flow. Time in h:min:sec.}
	\label{Mov_Bending_TempsLong}
\end{figure}

\subsection{Buckling}

\begin{figure}[H]
	\includegraphics[width=0.8\textwidth]{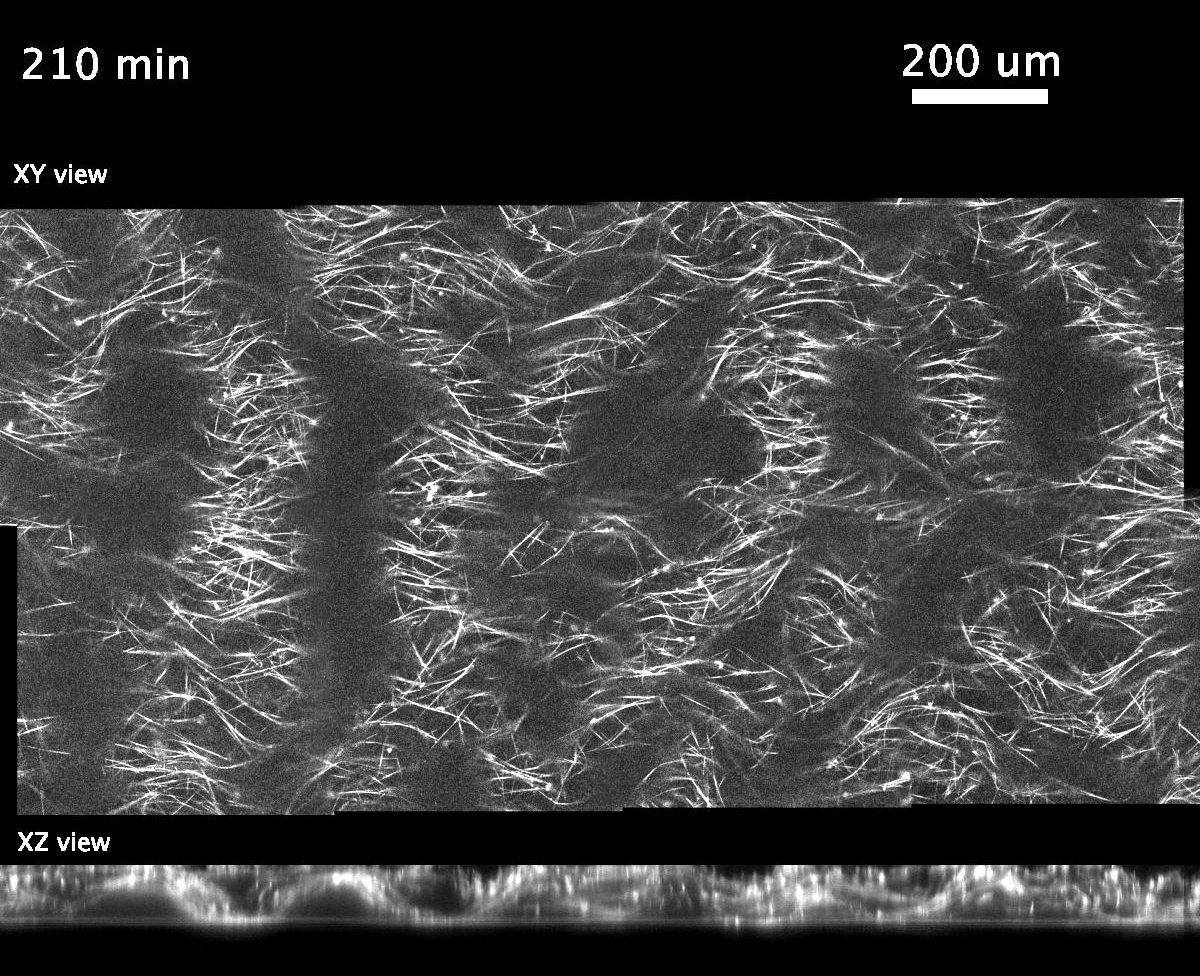}
	\renewcommand{\figurename}{Movie S\!\!}
	\caption{Time-lapse confocal microscopy of microtubules displaying a buckling instability. Associated to Figure 1 in the main text. 5.5\% pluronic, [ATP] = 5~$\mu$M.}
	\label{Mov_Fig1_Buckling}
\end{figure}

\begin{figure}[H]
	\includegraphics[width=0.5\textwidth]{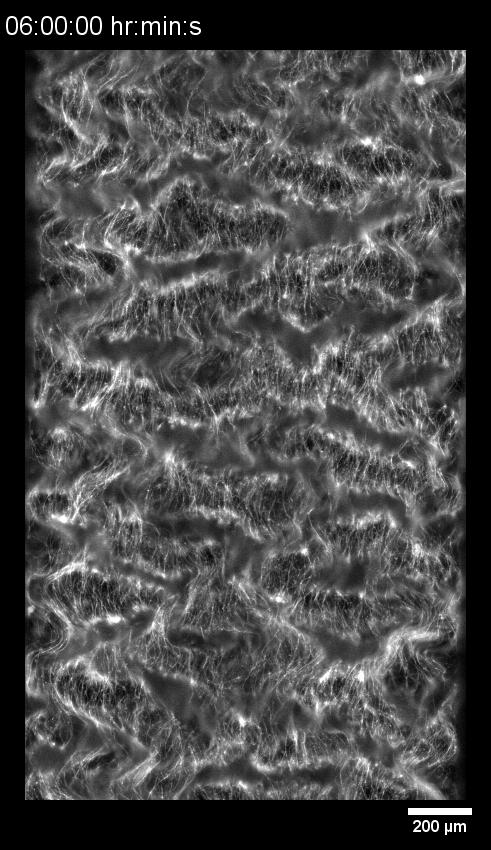}
	\renewcommand{\figurename}{Movie S\!\!}
	\caption{Time-lapse epifluorescence microscopy showing the long-time behavior of microtubules displaying a buckling instability. Within 1000 minutes, this instability turns into chaotic flow. Time in h:min:sec.}
	\label{Mov_Buckling_TempsLong}
\end{figure}

\subsection{Local contraction}

\begin{figure}[H]
	\includegraphics[width=0.8\textwidth]{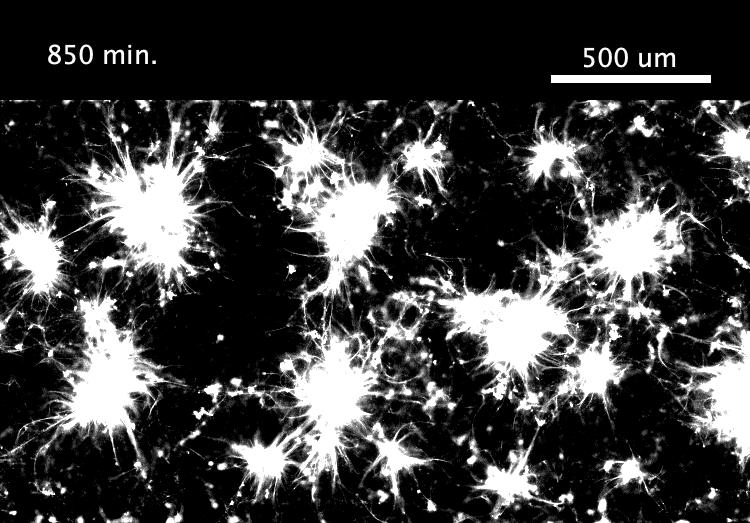}
	\renewcommand{\figurename}{Movie S\!\!}
	\caption{Time-lapse epifluorescence microscopy of microtubules displaying local contractions. Associated to Figure 2 in the main text. 1.5~\% pluronic, [ATP] = 100~$\mu$M.}
	\label{Mov_Fig1_LocalContraction}
\end{figure}

\subsection{Global contraction}

\begin{figure}[H]
	\includegraphics[width=0.8\textwidth]{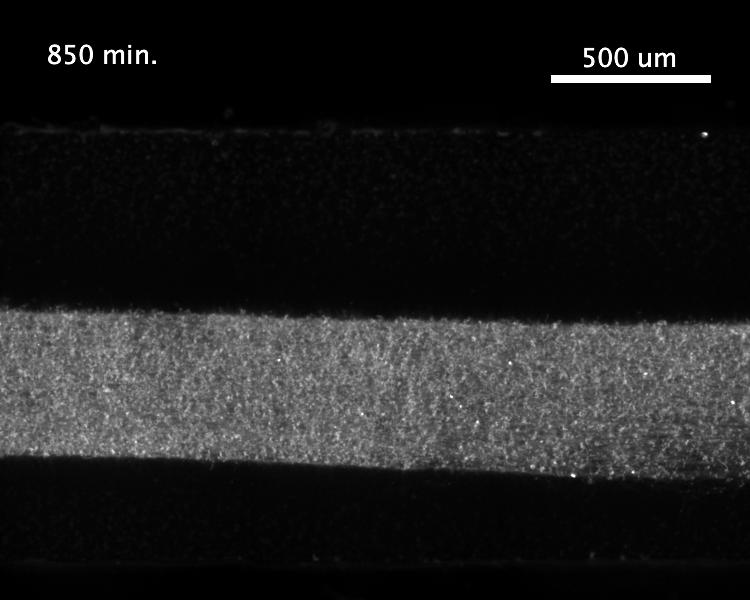}
	\renewcommand{\figurename}{Movie S\!\!}
	\caption{Time-lapse epifluorescence microscopy of microtubules displaying a global contraction. Associated to Figure 2 in the main text. 1.5~\% pluronic, [ATP] = 5~$\mu$M.}
	\label{Mov_Fig1_GlobalContraction}
\end{figure}

\subsection{Bending becomes buckling and local becomes global contraction with increasing PRC1}

\begin{figure}[H]
	\includegraphics[width=0.8\textwidth]{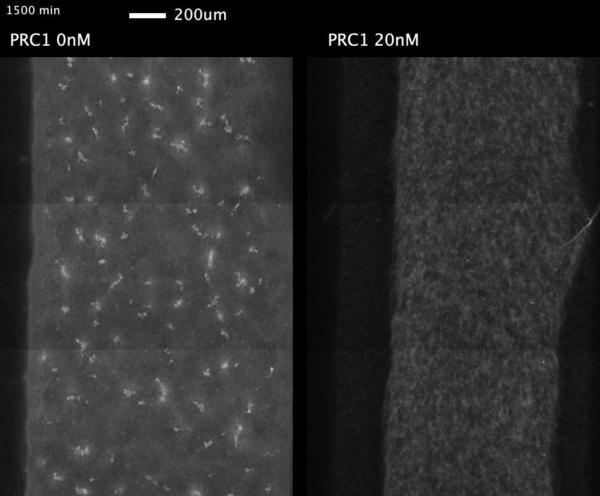}
	\renewcommand{\figurename}{Movie S\!\!}
	\caption{Time-lapse epifluorescence microscopy of microtubules displaying global or local contractions depending on the concentration of passive linkers at constant [ATP] and pluronic. 
		The addition of 20 nM passive linkers of PRC1 in a solution that displays local contractions (left) is sufficient to convert the local contractions into a global contraction (right). Associated to Figure 5 in the main text.}
	\label{Mov_PRC1_contraction}
\end{figure}

\begin{figure}[H]
	\includegraphics[width=0.8\textwidth]{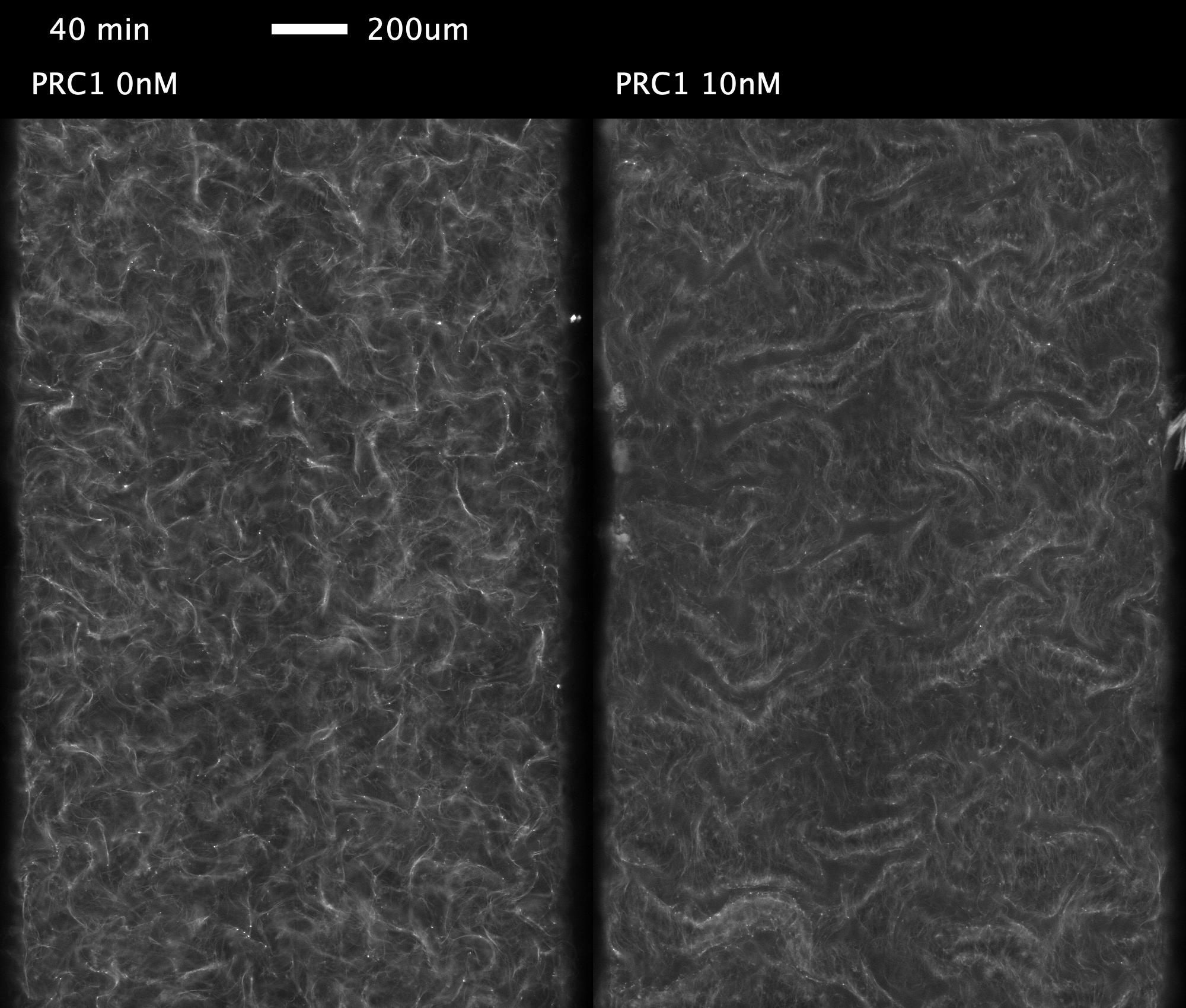}
	\renewcommand{\figurename}{Movie S\!\!}
	\caption{Time-lapse epifluorescence microscopy of microtubules displaying bending or buckling instabilities depending on the concentration of passive linkers at constant [ATP] and pluronic. 
		The addition of 10 nM passive linkers of PRC1 in a solution that displays a bending instability is sufficient to convert the bending instability into a buckling instability. Associated to Figure 5 in the main text.}
	\label{Mov_PRC1_bendingbuckling}
\end{figure}


\newpage
\providecommand{\latin}[1]{#1}
\makeatletter
\providecommand{\doi}
  {\begingroup\let\do\@makeother\dospecials
  \catcode`\{=1 \catcode`\}=2\doi@aux}
\providecommand{\doi@aux}[1]{\endgroup\texttt{#1}}
\makeatother
\providecommand*\mcitethebibliography{\thebibliography}
\csname @ifundefined\endcsname{endmcitethebibliography}
  {\let\endmcitethebibliography\endthebibliography}{}